\documentclass[12pt]{article}
\usepackage{amsmath,amsfonts, epsf,epsfig,color,latexsym,cite}
\usepackage{bbold}
\numberwithin{equation}{section}
\newcommand{\1}{\mathbb{1}}

\def\be{\begin{equation}}       \def\eq{\begin{equation}}
\def\ee{\label{abc}  \end{equation}}         \def\eqe{\label{abc}  \end{equation}}

\def\bea{\begin{eqnarray}}      \def\eqa{\begin{eqnarray}}
\def\ena{\end{eqnarray}}        \def\eea{\end{eqnarray}}
                                \def\eqae{\end{eqnarray}}

\def\a{\alpha}
\def\b{\beta}

\def\d{\delta}
\def\e{\epsilon}           
\def\f{\phi}               
\def\g{\gamma}
\def\h{\eta}
\def\i{\iota}

\def\k{\kappa}                    
\def\l{\lambda}

\def\p{\pi}                
  \def\th{\theta}                  
\def\r{\rho}                                     
\def\s{\sigma}                                   
\def\t{\tau}

\def\x{\xi}

\def\G{\Gamma}

\def\L{\Lambda}
  \def\W{\Omega}
\def\P{\Pi}
\def\Q{\Theta}
\def\SS{\Sigma}

\def\ca{{\cal A}}

\def\ce{{\cal E}}

\def\ch{{\cal H}}

\def\cj{{\cal J}}

\def\cm{{\cal H}}

\def\cv{{\cal V}}

\def\vx{v}
\def\vy{V}

\def\bop#1{\setbox0=\hbox{$#1M$}\mkern1.5mu
        \vbox{\hrule height0pt depth.04\ht0
        \hbox{\vrule width.04\ht0 height.9\ht0 \kern.9\ht0
        \vrule width.04\ht0}\hrule height.04\ht0}\mkern1.5mu}
\def\pa{\partial}                              
\def\we{\wedge}                                         
\def\>{\rangle} 

\def\<{\langle} 

\def\Tilde#1{\widetilde{#1}}                   

\def\to{\rightarrow}

\def\pa{\partial}

\def\ha{\frac12}                               

\def\CM{{\cal M}}

\def\IZ{\relax\ifmmode\mathchoice
{\hbox{\cmss Z\kern-.4em Z}}{\hbox{\cmss Z\kern-.4em Z}}
{\lower.9pt\hbox{\cmsss Z\kern-.4em Z}} {\lower1.2pt\hbox{\cmsss
Z\kern-.4em Z}}\else{\cmss Z\kern-.4em }\fi}
\def\IC{\relax\hbox{$\inbar\kern-.3em{\rm C}$}}
\def\IR{\relax{\rm I\kern-.18em R}}

\ifx\ltimes\undefined
\newcommand{\ltimes}{{\kern3pt\hbox{\vrule width 0.4pt height 5.30pt
depth .0pt}\kern-1.76pt\times\kern1pt}} \fi

\def\be{\begin{equation}}
\def\ee{\label{abc}  \end{equation}}
\def\ba{\begin{eqnarray}}
\def\ea{\end{eqnarray}}
\def\bq{\begin{quote}}
\def\eq{\end{quote}}

\def\part{\partial}

\def\beq{\begin{equation}}
\def\eeq{\label{abc}  \end{equation}}
\def\beqa{\begin{eqnarray}}
\def\eeqa{\end{eqnarray}}

\def\ti{\Tilde}

\def\Z {\mathbb{Z}}
\def\R {\mathbb{R}}
\def\C {\mathbb{C}}

\begin{document}
\thispagestyle{empty}
\begin{flushright}
hep-th/0701203\\
Imperial/TP/2007/CMH/01
\end{flushright}\vskip 0.8cm\begin{center}
\LARGE{\bf   Generalised Geometry for M-Theory}
\end{center}
\vskip 1in

\begin{center}{\large C M  Hull }
\vskip 0.2cm{ Theoretical Physics Group,  Blackett  Laboratory, \\
Imperial College,\\ London SW7 2BZ, U.K.}

\end{center}
\vskip 1.0cm

\begin{abstract}\noindent
Generalised geometry studies structures on a $d$-dimensional manifold  with a metric and 2-form gauge field  on which there is a natural action of the group $SO(d,d)$. This is generalised to $d$-dimensional manifolds with a metric and 3-form gauge field  on which there is a natural action of the group $E_{d}$. This provides a framework for the discussion of M-theory solutions with flux. A different generalisation is to $d$-dimensional manifolds  with a metric,  2-form gauge field and a set of $p$-forms for $p$ either odd or even on which there is a natural action of the group $E_{d+1}$. This is useful for type IIA or IIB string solutions with flux. Further generalisations give extended tangent bundles and extended spin bundles relevant for non-geometric backgrounds. Special structures that arise for supersymmetric backgrounds are  discussed.

\end{abstract}

\vfill

\setcounter{footnote}{0}
\def\thefootnote{\arabic{footnote}}
\newpage

\section{Introduction}\label{Introduction}

Hitchin's generalised geometry \cite{Hitchin:2004ut} -- \cite{Gualtieri}
studies structures on a $d$-dimensional manifold $M$ on which there is a natural action of the group $SO(d,d)$, and in particular it  gives an elegant description of geometries equipped with both a metric $G$ and a 2-form $B$. Such geometries with   a metric   and  2-form and an action of $SO(d,d)$ arise in string theory, so that this is a natural framework in which to formulate many problems in string theory and supergravity
\cite{Lindstrom:2004eh} 
--\cite{Bredthauer:2006hf}.
However, in type II string theory, the group  $SO(d,d)$ is part of a much larger \lq U-duality' group \cite{Cremmer:1979up},
\cite{Hull:1994ys}
$E_{d+1}$ that acts on $G$ and  $B$ together with a set of other fields on $M$ (the Ramond-Ramond gauge fields) and this suggests that seeking a generalisation of generalised geometry 
with $SO(d,d)$ replaced by $E_{d+1}$
might provide a natural framework for the geometries with flux in type II string theory.
Here $E_n$ is the maximally non-compact real form of the group  of rank-$n$ with   $E$-type Dynkin diagram, so that it is the
exceptional group $E_n$ for $n=6,7,8$. The U-duality groups $E_n$ and their maximal compact subgroups   $H_n$ 
are given in table 1 for $2\le n\le 8$. These groups were found to be symmetries of supergravity theories in
\cite{Cremmer:1979up}
and the global structure of these groups and their maximal subgroups was discussed in \cite{Keurentjes:2003hc},\cite{Keurentjes:2003yu}.

  M-theory has a similar structure in which there is a metric $G$ and 3-form $C$ on an $n$-dimensional manifold ${\cal M} $ with a  natural action of $E_{n}$, and again one might expect a generalisation of generalised geometry with the 3-form $C$ playing a central role. The relationship between M-theory and string theory suggests that if the manifold ${\cal M} $ is a circle bundle over a manifold $M$ of dimension $d=n-1$, then the M-geometry on ${\cal M} $ should reduce to a stringy generalised geometry on $M$.
  
  The aim of this paper is to propose such generalisations, and to set up the framework needed to study general supersymmetric string or M-theory backgrounds, including non-geometric ones. This will lead to the introduction of new structures, and in particular to extended tangent bundles and extended spin bundles for type II geometries and M-geometries. It will be convenient to refer to the usual generalised geometry involving $SO(d,d)$ as a type I geometry, to distinguish it from these other geometries, and it indeed plays a role in type I superstrings.

In generalised geometry, the tangent bundle $T$ is replaced with  $T\oplus T^*$, the sum of the tangent and cotangent bundles, which has a natural
inner product  of signature $(d,d)$ preserved by the  action of $SO(d,d)$. 
This group includes the action of a 2-form on the geometry, which acts as a shift of $B$.
A generalisation of the spin bundle is a bundle $S$ over $M$ with transition functions in $Spin(d,d)$.
Given a choice of spin structure,
there is a correspondence between  $S$  and  the bundle $\L ^\bullet T^*$ of formal sums of differential forms on $M$, and $S$ splits into the chiral and anti-chiral sub-bundles
$S^+$ and $S^-$ corresponding to even and odd forms respectively.
The perturbative charges of string theory (momentum and string charge) fit into a vector of $SO(d,d)$.
In addition, there are Ramond-Ramond charges which are even forms for the type IIA string and odd forms for the IIB string, and  the Ramond-Ramond charges transform according to the spinor representation of the $SO(d,d)$ subgroup of the U-duality group \cite{Hull:1994ys}.
This suggests that $T\oplus T^*$ be extended to $T\oplus T^*\oplus S^+$  for IIA or $T\oplus T^*\oplus S^-$ for IIB. 
This turns out to be sufficient for $d\le 4$, but for $d>4$ there are further charges consisting of a five-brane charge given by a 5-form in $\L^5 T^*$ and a charge related to the Kaluza-Kein monopoles
\footnote{In $D=10$ or $D=11$, there is a 5-form charge in the superalgebra,
$Z_{M_1...M_5}$. Decomposing the indices $M=(0,i)$ where $i=1,...,D-1$ is a spatial index and $0$ is a time index gives two charges, a spatial 5-form charge
$Z_{i_1...i_5}$ which is the NS-NS or M-theory 5-brane charge, and a spatial 4-form charge $Z_{0i_1...i_4}$, which is the Kaluza-Kein monopole charge, Hodge-dual to a spatial $D-5$-vector
$Z^{i_1...i_{D-5}}$  \cite{Hull:1997kt}. This gives charges in
$\L ^5  T\oplus \L ^5  T^*$ for $D=10$ or in $\L ^6  T\oplus \L ^5  T^*$
for $D=11$.}
  \cite{Hull:1997kt} represented by a 5-vector in $\L^5 T$, so that for type II strings the tangent bundle is generalised to
 the extended tangent bundle
 $$T\oplus T^*\oplus\L^5 T \oplus \L^5 T ^*\oplus S^\pm$$
 As will be seen in section \ref{type2}, there is a natural action of $E_{d+1}$ on this space for $d\le 6$.

 A bundle with structure group $O(d,d)$ is reducible to an $O(d)\times O(d)$ bundle. In generalised geometry, the metric $G$ and 2-form $B$ arise as the moduli for such   reductions, and parameterise a coset space
  $O(d,d)/O(d)\times O(d)$. This is generalised to the coset $E_{d+1}/H_{d+1}$ 
  which can be parameterised by a  metric $G$ and 2-form $B$ and scalar $\Phi$, together with a set of odd forms
  $C_1,C_3,...$ for IIA geometries or a set of even forms $C_0,C_2,C_4,...$ for IIB geometries.
  These extra fields have a natural interpretation in type II string theory as the dilaton $\Phi$ and the Ramond-Ramond $p$-form gauge fields $C_p$. The formal sums  $C^+=C_0+C_2+C_4+...$ or $C^-=C_1+C_3+C_5+...$
  transform as
   chiral spinors under $Spin(d,d)$, with   the index $\pm$ indicating the chirality.  
 The  action of $E_{d+1}$ on these fields includes shifts for each of  the $p$-form gauge fields of the theory.

 Comparison with M-theory suggests 
a different generalisation,  replacing  $T^*$ (corresponding to a string charge) with $\L^2T^*$
 (corresponding to a membrane charge), so that
 the extended tangent bundle includes    $T\oplus \L^2 T^*$. For manifolds of dimension $n>4$, it is necessary to add  $\L^5T^*$  (corresponding to a 5-brane charge), and for $n>5$ an additional 
$\L^6T$  (the Kaluza-Kein monopole charge  \cite{Hull:1997kt}) is needed. Then for $n\le 7$, the 
extended tangent bundle is  
$$T\oplus \L^2 T^*\oplus \L^5 T^*\oplus \L^6 T$$
There is a natural action of $E_{n}$ on this.
The coset space $E_{n}/H_{n}$  can be  parameterised by 
a metric $G$, a 3-form   $C$ and (for $n\ge 6$) a 6-form   $\ti C$ on the $n$-dimensional manifold.
The 3-form $C$ can be associated with the 3-form gauge field of 11-dimensional supergravity, and 
the 6-form   $\ti C$  with the dual gauge field. (Recall that a free 3-form gauge field in 11-dimensions 
has a dual representation in terms of a 6-form gauge field, related by an electromagnetic duality, $d\ti C_6 \sim *dC_3$.
The Chern-Simons interaction of 11-dimensional supergravity prevents the dualisation  to a theory written in terms of a 6-form gauge field, but it can be written in terms of both a 3-form $C$ and a 6-form $\ti C$, \cite{Cremmer:1998px}.)
The  action of $E_{n}$ on these fields includes shifts of the 3-form   field $C$ and   6-form   field $\ti C$.

 \begin{table}
\begin{center}
\begin{tabular}{|c|cccc|}
\hline
n & $E_n$ & $H_n$ & dim($E_n$)& dim($E_n/H_n$) \\
\hline
2  & $SL(2,\R) \times \R$ & $SO(2)$ & 4& 3\\
3  & $SL(3,\R) \times SL(2,\R)$ & $SO(3) \times SO(2)$& 11&7 \\
4  & $SL(5,\R)$ & $SO(5)$ & 24& 14\\
5  & $Spin(5,5)$ & $(Sp(2) \times Sp(2))/\Z_2$&  45& 25\\
6  & $E_{6(6)}$ & $Sp(4)/\Z_2$ & 78&42\\
7  & $E_{7(7)}$ & $SU(8)/\Z_2$ & 133 & 70\\
8  & $E_{8(8)}$ & $Spin(16)/\Z_2$ & 248 & 128\\
\hline    
\end{tabular}
\caption{The U-duality groups $E_n$, their maximal compact
  subgroups $H_n$,  and the     dimensions of $E_n$ and the cosets $E_n/H_n$.} \label{tab1} 
\end{center}
\end{table}

For a $d$-dimensional manifold, the structure group of $T$, $T^*$, $T\oplus  T^*$ (and their tensor products) is  in $GL(d,\R)$, which is a subgroup of $O(d,d)$. Twisting with a gerbe can enlarge the structure group to include  the action of exact 2-forms \cite{Hitchin:2004ut},\cite{Hitchin:2005in},\cite{Gualtieri}, but this is still only a   part of $O(d,d)$; this is the \lq geometric subgroup'  that preserves the Courant bracket. However, the covariance under the larger group $O(d,d)$ is very suggestive, and this suggests that bundles with this larger structure group might have an interesting role to play. String theory can in fact be formulated on a large class of spaces with so-called non-geometric structures, and 
including these allows a wider class of transition functions.
For example, for string theory on a manifold $M$ that is an $m$-torus bundle with fibres  ${\bf T}^m$,
there is a symmetry under the action of the T-duality group $O(m,m;\Z)$, which in particular mixes the metric and $B$-field together.
This symmetry allows the construction of  T-folds.  These are spaces  built  from patches which are each of the form $U_\a\times {\bf T}^m$   with $U_\a$   open sets in the base, and  with transition functions that include  $O(m,m;\Z)$    T-duality transformations \cite{Hull:2004in}.
As the patching is through symmetries of the theory, it leads to consistent backgrounds of string theory.
However,  these are not manifolds equipped with tensor fields but are considerably more general.
The generalised tangent bundle for such spaces has $O(d,d)$ transition functions not contained within the geometric subgroup.
These have generalisations to U-folds with fibres  ${\bf T}^m$ whose transistion functions include transformations in the U-duality group $E_{m+1}(\Z)$ \cite{Hull:2004in}, and the extended geometries discussed here provide a natural framework to discuss these geometries. Examples of T-folds have been studied in \cite{Dabholkar:2002sy}
-\cite{Gray:2005ea}.

\section{Generalised Geometry}\label{Geometry}

\subsection{The Structure  of Generalised Geometry} \label{General}

In Hitchin's generalised geometry, the tangent bundle $T$ of a $d$-dimensional manifold $M$
is replaced with  $T\oplus T^*$, so that one considers the formal sum $V=\vx +\xi$ of a vector field $\vx $ with components $\vx ^i$ ($i=1,...,d$) and a one-form $\xi$ with components $\xi_i$, which can be thought of as a 
vector with $2d$ components $V^I$
\begin{equation}
V^I=\left(\begin{array}{cc}
\vx ^i   \\
\xi_i
\end{array}\right),
\label{vec}
 \end{equation}
There is a natural inner product $\eta$ of signature $(d,d)$ defined by
$$\eta (\vx +\xi,\vx +\xi)=2\i_\vx \xi$$
where $\i$ denotes the interior product, so that   $\i_\vx \xi = \vx ^i\x_i$.
The metric has components $\eta _{IJ}$ given by
\begin{equation}
\eta=\left(\begin{array}{cc}
0 & \1  \\
\1 & 0
\end{array}\right),
\label{dsfsdh}
 \end{equation}
 This is invariant under the orthogonal group $O(d,d)$, with $V$ transforming in the vector representation
 $V \to gV$, where $g$ is represented by a matrix $g^I{}_J$ satisfying
 \begin{equation}
g^t\eta g=\eta
\end{equation}
The Lie algebra of $O(d,d)$ consists of matrices with the block decomposition
\begin{equation}\label{comp}
\left(\begin{matrix}A&\beta\\ \Theta&-A^t\end{matrix}\right),
\end{equation}
Here $A^i{}_j$ is an arbitrary   $d\times d$ matrix, and so is a generator of the  $GL(d,\R)$   subgroup of 
matrices $g$ of the form
\begin{equation}
\left(\begin{matrix}M&0\\0&(
M^t)^{-1}\end{matrix}\right),\label{gldr}
\end{equation}
for arbitrary invertible matrices $M^i{}_j$.
The $\Theta_{ij}$ are components of a 2-form $\Theta\in \Lambda^2T^*$ generating the group of matrices
\begin{equation}\label{thgrp}
\left(\begin{matrix}\1&0\\ \Theta&\1 \end{matrix}\right),
\end{equation}
sending 
\begin{equation}
\label{bshift}
\vx +\xi\mapsto \vx +\xi+\i_\vx \Theta
\end{equation}
while $\b\in \L^2 T$ is a generator of the group of matrices of the form
\begin{equation}
\left(\begin{matrix}\1&\beta\\0&\1\end{matrix}\right),
\end{equation}
 sending 
\begin{equation}
\label{beshift}
\vx +\xi\mapsto
\vx +\xi+\i_\xi\beta
\end{equation}
The \lq geometric subgroup' $GL(d,\R)\ltimes \R^{d(d-1)/2}$ generated by $A,\Q$  of matrices of the form
\begin{equation} \label{geomgroup}
\left(\begin{matrix}M&0\\\Q &(
M^t)^{-1}\end{matrix}\right),
\end{equation}
will play a role in what follows.

There is a natural action of $Spin(d,d) $ on  the bundle of  formal sums of differential forms $\L ^\bullet T^*$ on $M$, so that interesting geometric structures can be formulated in terms of spinors. For each $V=\vx+\x \in T\oplus T^*$, there is a map
$\G _V: \L ^\bullet T^* \to \L ^\bullet T^*$ such that 
$$ \G_V:  \f \mapsto  \i_\vx \f + \x \wedge \f$$
for any $\f\in  \L ^\bullet T^*$.
These maps satisfy a Clifford algebra, with
\begin{equation}
\label{}
\G_V \G_{V'} + \G_{V'}\G_V= -2 \h(V,V') \1
\end{equation}
The Clifford action on 
 $\L ^\bullet T^*$ gives in particular a representation of $Spin(d,d) $ on  $\L ^\bullet T^*$.
 The action of $GL(d,\R)\subset Spin(d,d)$ on $\L ^\bullet T^*$  is not quite the usual one.
 If the standard action of  $M\in GL(d,\R)$ on $\L ^\bullet T^*$ is denoted $M^*$, the action of 
$GL(d,\R)\subset Spin(d,d)$
is $$ \f \mapsto | {\rm det}~ M |^{1/2} M^* \f$$
so that the relation with the  spin bundle $S$ is 
$$S= \L ^\bullet T^* \otimes (\L ^d T)^{1/2}$$
The bundle of forms splits into the bundle $\L ^+ T^*$ of even forms and the bundle $\L ^- T^*$ of odd forms, corresponding to the decomposition of $S$ into bundles $S^\pm$ of positive or negative chirality spinors, with
$$S^\pm= \L ^\pm T^* \otimes (\L ^d T)^{1/2}$$
The bundle $(\L ^d T)^{1/2}$ is trivial and so there is always a non-canonical isomorphism $S^\pm \sim \L^\pm T^*$; $S^\pm $ and $ \L^\pm T^*$ will be used interchangably for the remainder of the paper.
(There is in addition another possible spin structure \cite{Gualtieri}, but this will not be used here.)

The Courant bracket provides a generalisation of the Lie bracket to $T\oplus T^*$, and plays a central role in generalised geometry, and is preserved under (\ref{bshift}) provided
 $\Theta$ is closed.
According to Hitchin \cite{Hitchin:2005in}, generalised geometries are structures on $T\oplus T^*$ that are compatible with
 the $SO(d,d)$  structure and which satisfy integrability conditions expressed in terms of the Courant bracket or the exterior derivative.

The transition functions for $M$ are diffeomorphisms, so that the  transition functions for 
$T\oplus T^*$ are in $GL(d,\R)$, although it is sometimes useful to instead regard it as having structure group in $SO(d,d)$ \cite{Gualtieri}. This can be generalised by twisting with a gerbe, as will be reviewed in the next subsection.
For $d=2m$, a generalised almost  complex structure is an endomorphism ${\cal J}$ of   $T\oplus T^*$ that satisfies
${\cal J}^2=-\1$ and with respect to which the metric $\h$ is hermitian. It is a generalised    complex structure
if it is integrable, i.e. the $+i$-eigenbundle $E <   (T\oplus T^*)\otimes \C$ is
such that the space of sections of $E$ is closed under the Courant bracket.
Such a structure is preserved under the $U(m,m)$ subgroup of $ SO(2m,2m)$.

Gualtieri introduced the concept of a generalised metric $\cm$ on $T\oplus T^*$ \cite{Gualtieri}.  This  is  a positive definite
metric compatible with $\eta$, and defines a   sub-bundle $E_+$ 
on which $\h$
 is positive definite.
The generalised metric  can be  represented by a matrix $\cm_{IJ}$ satisfying the compatibility condition
\begin{equation}
\eta ^{-1} \cm \eta^{-1}=\cm^{-1}
\label{gmetc}
\end{equation}
This implies that $S^I{}_J$ defined by
\begin{equation}
S= \eta^{-1} \cm
\end{equation}
satisfies
\begin{equation}
S^2=\1
\end{equation}
and so is an almost real structure or almost local product structure.
($S$ is sometimes also referred to as the generalised metric \cite{Gualtieri}.)
It has $d$ eigenvalues of $+1$ and $d$ eigenvalues of $-1$, and $E_+$ is the $+1$ eigenbundle.

The constraint (\ref{gmetc}) implies that $\cm$ has $d^2$ independent components and it can be parameterised in terms of a symmetric matrix $G_{ij}$ and an anti-symmetric matrix $B_{ij}$ as
\begin{equation}
\cm  =\left(
\begin{array}{cc}
G-BG^{-1}B & BG^{-1} \\
-G^{-1}B   & G^{-1}
\end{array}
\right)
\label{hiss}  \end{equation}
and   $\cm$ is positive definite if $G$ is. 
The norm of the vector $V=\vx+\x$   is then
\begin{equation}
\label{}
\cm(V,V)=G(\vx ,\vx )+ G^*(\x+\i_\vx B,\x+\i_\vx B)
\end{equation}
where $G^*$ is the metric on $T^*$ given by the inverse of $G$ and $(\i_\vx B)_i= v^jB_{ji}$.
Thus introducing a generalised metric is equivalent to introducing a positive definite metric $G$  and a  2-form $B$ on $M$.
This can be generalised to   a metric $G$ of signature $(p,q)$ on $M$, in which case the generalised metric given by (\ref{hiss}) has signature $(2p,2q)$.

Under an $SO(d,d)$ transformation
\begin{equation}
\cm \to g^t \cm g
\label{gtrans}  \end{equation}
This corresponds to a fractional linear transformation of $G,B$. Defining the $d\times d $ matrix
\begin{equation}
E_{ij}=G_{ij}+B_{ij}
\label{abc}  \end{equation}
and decomposing $g$ into $d\times d$ matrices  $a,b,c,d$
 \begin{equation}
g=\left(
\begin{array}{cc}
a & b\\
c & d \end{array}\right)
\label{abc}  \end{equation}
 so that 
 \begin{equation}
g^t\eta g=\eta \;\; \Rightarrow \;\;a^t c+c^t a=0,\;\;\;b^t d+d^t
b=0,\;\;\;\ a^t d+c^t b=\1  , \label{abc}  \end{equation}
then the transformation of $G,B$ under the action of the $SO(d,d)$ transformation
$g$ is
\begin{equation}
E'  = (aE+b)(cE+d)^{-1}. \label{tetrans}  \end{equation} 
In particular, the action of the $GL(d,\R)$ subgroup (\ref{gldr}) is the linear transformation
\begin{equation}
G\to M^t GM, \qquad B\to M^t BM,
\end{equation}
while the $\Q$ transformation (\ref{thgrp}) leaves $G$ invariant and  acts as a shift of $B$:
\begin{equation}
B\to B+\Q
\end{equation}
However,   $SO(d,d)$ transformations not in the geometric subgroup will mix $G $ and $B$.

\subsection{Gerbes and the Generalised Tangent Bundle}\label{gerbe}

For $T\oplus T^*$,  the structure group is $GL(d,\R)$ and introducing a generalised metric corresponds to introducing a symmetric tensor field $G$ and an anti-symmetric tensor field $B$ on $M$.
However, this can be generalised to allow $B$ to be a gerbe conection, i.e. a 2-form gauge field with field strength $H=dB$, allowing a twisting of this construction to allow transition functions including the $B$-shift.

Given an open cover $\{U_{\alpha}\}$ of  $M$, there is a 2-form $B_\a$ in each
$\{U_{\alpha}\}$ with $B_\beta-B_\alpha$ a closed 2-form on the overlap $U_{\alpha}\cap U_{\beta}$, so that  $dB_{\beta}=dB_\alpha=H$ is a globally defined closed three-form $H$. For a suitable open cover, the 
overlaps   have trivial cohomology and
$$B_\beta-B_\alpha=d\l_{\alpha\beta}$$ 
for some 1-form $\l_{\alpha\beta}$ on the overlap $U_{\alpha}\cap U_{\beta}$.
 Consistency on triple overlaps $U_{\alpha}\cap U_{\beta}\cap U_{\gamma}$ 
 requires that $\l_{\alpha\beta}+\l_{\beta\gamma}+\l_{\gamma\alpha}$ is  closed and so exact. If  it is of the form
 $$\l_{\alpha\beta}+\l_{\beta\gamma}+\l_{\gamma\alpha}= g_{\alpha\beta\gamma}^{-1}dg_{\alpha\beta\gamma}$$
 for some $U(1)$-valued functions 
 $$g_{\alpha\beta\gamma}:U_\alpha\cap U_\beta\cap U_\gamma\rightarrow S^1$$
satisfying  $g_{\alpha\beta\gamma}=g^{-1}_{\beta\alpha\gamma} $ and 
 $g_{\beta\gamma\delta}g^{-1}_{\alpha\gamma\delta}g_{\alpha\beta\delta}g^{-1}_{\alpha\beta\gamma}=1$ 
   on $U_\alpha\cap U_\beta\cap U_\gamma \cap U_\delta$, then $B_\a$ defines a connection on a gerbe and $H$ represents an integral cohomology class.
 (If $H$ is not in an integral cohomology class, then  
 $$\l_{\alpha\beta}+\l_{\beta\gamma}+\l_{\gamma\alpha}= d\rho _{\alpha\beta\gamma}$$
   for some 0-form  $\rho _{\alpha\beta\gamma}$ in  $U_{\alpha}\cap U_{\beta}\cap U_{\gamma}$
   satisfying a further consistency condition in quadruple overlaps.)
   
The    $\l_{\alpha\beta}$ can be used to define a bundle $E$ over $M$
by identifying $T\oplus T^*$ on $U_{\alpha}$ with $T\oplus T^*$ on $U_{\beta}$ by the B-field action $$\vx+\xi\mapsto \vx +\xi+\i_\vx d\l_{\alpha\beta}$$
  The fibre over a point $x$ in $M$ is again
$T_x\oplus T^*_x$, but   the transition functions are no longer in $GL(d,\R)$.
The bundle $E$ has been called a generalised tangent bundle \cite{Hitchin:2005in} and has a structure group in the geometric subgroup of $SO(d,d)$, i.e. the subgroup
$GL(d,\R)\ltimes \W^{2, cl}$, where $ \W^{2, cl}$ is  the space of closed 2-forms.

 \section{The Structure of Extended Geometries}\label{struc}
 
\subsection{Type  I Extended Geometries: Generalising the Generalised Tangent Bundle and Spin Bundle}\label{gengen}

To incorporate   structures such as T-folds and other  non-geometric backgrounds, it is useful to generalise the structure further and consider general bundles $E$ over a $d$-dimensional space $M$ with structure group
$O(d,d)$ or $SO(d,d)$ and split-signature fibre metric $\h$; these will be generalised geometries in the sense of Hitchin  only in the special case in which the structure group is in the geometric subgroup preserving the Courant bracket, and will only correspond to  $T\oplus T^*$ if the structure group is in $GL(d,\R)$.
Locally, one can find a metric $G$ and 2-form $B$ as before, but general $O(d,d)$ transition functions mix $G$ and $B$, so that these will not be  tensor fields on $M$ in general, and the background will be \lq non-geometric'. Nonetheless, such backgrounds with $m$-torus fibrations and transition functions including $O(m,m;\Z)$ transformations arise in string theory as T-folds, so that this is a useful generalisation. 
Such extended geometries with  $O(d,d)$  structure
will be referred to as Type I extended geometries, to distinguish them from the type II and type M geometries with E-series structure groups to be introduced later.
It will also be natural to introduce an extended spin bundle $S$ with structure group
$Pin(d,d)$ or $Spin(d,d)$, when there is no obstruction to such a double cover of $E$.

The bundle $E$ can be reduced to one that has  structure group in the maximal compact subgroup $O(d)\times O(d)$ or
$S(O(d)\times O(d))$. This is equivalent to choosing a sub-bundle $E^+$ on which $\h$ is positive definite, so that $E=E^+\oplus E^-$ where $E^-$ is the orthogonal complement of $E^+$, so that $\h$ is negative definite on $E^-$. An $SO(d,d)$  bundle $E$ admits a $Spin(d,d)$ structure only if the second Stiefel-Whitney classes of $E^\pm$ agree, $w_2(E^+)=w_2(E^-)$ \cite{Gualtieri},\cite{Karoubi}; this is automatically satisfied for $T\oplus T^*$, even in the case in which $M$ is not spin, i.e.  even if  $w_2(T)\ne 0$.

The reduction of $E$ to $E^\pm$ defines a positive definite generalised  metric
\begin{equation}
\label{}
\cm = \h\vert_{E^+}- \h\vert_{E^-}
\end{equation}
 Choosing a generalised  metric is equivalent to choosing a reduction
of the bundle, and the space of such reductions at a point $x\in M$ is
\begin{equation}
\label{}
\frac {O(d,d)} {O(d)\times O(d)}  ~~{\rm or }~~\frac {SO(d,d)} {S(O(d)\times SO(d))}
\end{equation}
Let $\cv ^\pm$ be the projections $\cv ^\pm : E \to E^\pm$.
Then
\begin{equation}
\cv  =\left( \begin{array}{c} \cv  ^+\\ \cv -\end{array}\right)
\label{abc}  \end{equation}
maps $E\to E^+\oplus E^-$
and is a representative of the coset $ {O(d,d)} /{O(d)\times O(d)}$.
Introducing indices $a=1,...,d$ labelling a basis for  $E^+$ transforming under one $O(d)$ factor and
indices $a'=1,...,d$ labelling a basis for  $E^-$ transforming under the other $ O(d)$ factor, $\cv^+$ is represented by a $d\times 2d$ matrix $\cv ^a{}_I$ and 
$\cv^-$ is represented by a $d\times 2d$ matrix $\cv ^{a'}{}_I$, so that
\begin{equation}
\cv ^A {}_I =\left( \begin{array}{c} \cv  ^a {}_I \\ \cv ^{a'} {}_I \end{array}\right),
\label{abc}  \end{equation}
is a vielbein  transforming from a general basis labelled by $I$ to a basis for $E^+\oplus E^-$ labelled by $A=(a,a')$. The generalised metric is then
\begin{equation}
\label{}
\cm= \cv ^t \cv
\end{equation}
with components
\begin{equation}
\label{}
\cm_{IJ}= \d_{AB}\cv^A{}_I \cv^B{}_J
\end{equation}
The generalised metric is not constant over $M$ in general, so $\cm(x)$ (where $x\in M$)
defines a map $\cm:M\to  { O(d,d)} /{O(d)\times O(d)}$. 
As well as the manifest covariance under $O(d,d)$, there is a symmetry under local $ {O(d)\times O(d)}$
transformations, given by functions $k(x)$, with $k:M\to  {O(d)\times O(d)}$.
In particular,  the vielbein $\cv(x)$ transforms as
\begin{equation}
\cv (x) \to k(x) \cv(x) g \label{abc}  \end{equation}
under a local  $ {O(d)\times O(d)}$ transformation $k(x)$ and rigid transformation $g\in O(d,d)$.
The local $O(d)\times O(d)$ symmetry
can be used to choose a triangular gauge for $\cv$ over some neighbourghood of $M$, so that
\begin{equation}
\cv= \left(\begin{array}{cc} e^t & 0 \\  -e^{-1}B
&e^{-1}\end{array}\right), \label{vist}  \end{equation} for some
$d$-bein $e_i{}^a$ and anti-symmetric $d\times d$ matrix $B_{ij}$.
Then
\begin{equation}
\cm= \cv ^t \cv =\left(
\begin{array}{cc}
G-BG^{-1}B & BG^{-1} \\
-G^{-1}B   & G^{-1}
\end{array}
\right).
\label{abc}  \end{equation}
where
the metric $G=e^te$, i.e.
\begin{equation}
G_{ij}= e_i{}^a e_j{}^b \d _{ab}
\label{abc}  \end{equation}
As a result, the fibre metric $\cm (x) $ is parameterized by
a $d\times d$ matrix $E(x)$ given by
\begin{equation}
E_{ij}=G_{ij}+B_{ij}
\label{abc}  \end{equation}

\subsection{General Extended Geometries}\label{GenEx}

The above structure generalises to  arbitrary vector bundles with non-compact structure group $G$.
Consider a vector bundle ${\cal E}$ over a manifold $M$ with projection $\p:{\cal E} \to M$, fibre $F$ and structure group $G$.
For an   open cover $\{ U_\a \}$ of $M$, $\p ^{-1} (U_\a) \sim U_\a\times F$ and a point in $\p ^{-1} (U_\a)$ 
can be represented by $(x_\a, \vy _\a)$ where
$x_\a\in U_\a$, $  \vy _\a \in F$. 
The group $G$ acts as $(x, \vy )\to (x,g \vy )$, where $g \vy \equiv R(g) \vy $ and
 $R(g)$ is the action of $g\in G$ on $F$ in some representation $R$.
Over the overlap $U_\a\cap U_\b$, the coordinates in $\p^{-1}(U_\a\cap U_\b)$ are related by
\begin{equation}
\label{vtrans}
 \vy _\a = g_{\a\b}(x) \vy _\b
\end{equation}
where the transition function $g_{\a\b}(x)$ is a map $g_{\a\b}:U_\a\cap U_\b\to G$
acting on $F$   (and satisfying the usual consistency conditions).

For any maps $\cv_\a: U_\a\to G$, the transition functions
\begin{equation}
\label{hist}
h_{\a\b} = \cv _\a g_{\a\b}  \cv _\b^{-1}
\end{equation}
define a bundle equivalent to $\ce$.
If $G$ is non-compact with maximal compact subgroup $H$, then 
${\cal E} $ can be reduced to a bundle $\bar {\cal E} $ with structure group $H$.
This means that the maps  $\cv_\a: U_\a\to G$ can be chosen so that
the transition functions (\ref{hist}) are in $H$, $h_{\a\b}\in H$.
For any such maps $\cv_\a$, the maps  $\cv '_\a= h_\a \cv_\a$ will also give transition functions in $H$, provided that $h_\a$ are maps $h_\a: U_\a \to H$. Then a reduction corresponds to an equivalence class of maps $\cv_\a$ identified under the left action of maps $h_\a: U_\a \to H$, $\cv _\a\sim  h_\a \cv_\a$. The equivalence classes   then correspond to maps from  $U_\a$ to the left coset $G/H$.
From (\ref{hist}), the maps $\cv_\a$ have the patching conditions
\begin{equation}
\label{}
\cv_\a = h_{\a\b} \cv_\b g_{\a\b}^{-1}
\end{equation}

There is then a map
$$\cv : {\cal E} \to \bar {\cal E}, \qquad \cv :(x_\a,  \vy _\a) \to (x_\a, \bar  \vy _\a)\equiv (x_\a,  \cv_\a (x_\a)) $$
where the $ \bar  \vy _\a = \cv_\a (x_\a) \vy _\a$ have patching conditions at $x$
\begin{equation}
\label{vtransb}
\bar  \vy _\a = h_{\a\b}(x)\bar  \vy _\b
\end{equation}
with transition functions $h_{\a\b}\in H$, so that 
$ \bar {\cal E}$ is indeed a vector bundle with structure group $H$.

Suppose that  the representation $R$  has an $H$-invariant  positive definite metric, giving a positive definite fibre metric $\bar \cm(\bar s, \bar s)$ for sections $\bar s(x)$ of $\bar \ce$,
and this in turn defines a positive definite fibre metric for sections $  s(x)$ of $  \ce$, via
\begin{equation}
\label{mett}
\cm(s,s)=\bar \cm(\cv s, \cv s)
\end{equation}
For example, if $H$ is an orthogonal group with $h^th=\1$ where $h^t$ is the transpose, then
$\bar \cm(\cv s, \cv s)= \bar s ^t \bar s$ and $ \cm( s,  s)=   s ^t \cm s$
where the matrix $\cm$ is given by
\begin{equation}
\label{}
\cm = \cv^t \cv
\end{equation}
For  $G=O(d,d)$, this gives the $O(d)\times O(d)$ invariant metric (\ref{hiss}).
Similarly, for unitary groups with $h^\dagger  h=\1$,
\begin{equation}
\label{}
\cm = \cv^\dagger \cv
\end{equation}

There is a natural action of $H$ gauge transformations, i.e. of maps $h_\a: U_\a\to H$ under which
\begin{equation}
\label{}
\cv_\a(x) \to h_\a(x) \cv_\a, \qquad (x,\bar  \vy _\a) \to (x,h_\a(x) \bar  \vy _\a), \qquad h_{\a\b} \to h_\a h_{\a\b} h^{-1}_\b
\end{equation}
We will be interested in gauge equivalence classes identified under this action. In particular, the metric $\cm$ depends only on the equivalence class, and so
is specified by a map $ M\to G/H$, or more generally a section of a bundle with fibre $G/H$.

Finally, for many cases of interest, $H$ has a natural double cover $\ti H$, and so given the extended tangent bundle 
$\bar \ce$ with $H$-structure, it is natural to seek an extended spin-bundle $\ti \ce$ with
structure group $\ti H$ that projects onto $\bar \ce$ under the double cover map $p:\ti H \to H$. There is
in general a topological obstruction for such a double cover, given by the 2nd Stiefel-Whitney class 
$w_2(\bar \ce) =H^2(\bar \ce , \Z_2)$.
Given a lift of the transition functions $h_{\a\b}$ to $\ti h_{\a\b}\in \ti H$, 
the $\Z_2$ Cech cohomology class is represented by the $\ti h_{\a\b} \ti h_{\b\g}\ti h_{\g\a}= \pm \1$ in triple overlaps, and it is necessary to be able to choose the $\ti h_{\a\b}$ so that this is $+\1$ in all triple overlaps.
A necessary and sufficient condition for this is that $w_2(\bar \ce) =0$.

In the following sections, examples of this construction with $G=E_n$ and $H=H_n$ will be explored.

\section{M-Geometries}

In this section, the generalisation of  generalised geometry suggested by M-theory
on an orientible $n$-dimensional manifold ${\cal M}$
are investigated, 
in which $T\oplus T^*$ with a natural action of $SO(n,n)$ is replaced by $\ce\sim T\oplus \L^2T^*  \oplus \dots $ with a natural action of $E_n$,
and the 2-form symmetry of $B$-shifts is generalised to one of 3-form shifts.
The structure changes from dimension to dimension, so each will be considered in turn.
The full  explicit transformations will be given only for $n=4,7$; those for $n=5,6$ follow by truncation of the $n=7$ case.

\subsection{$n=4$, $E_4=SL(5,\R)$}

Consider first the case of a four manifold, with $E_4=SL(5,\R)$. 
The bundle  $T\oplus T^*$ is replaced with $ T\oplus \L^2T^* $ with 10-dimensional fibres 
transforming in the $4+ 6$ representation of $SL(4,\R)$.
 A section is then a formal sum
 $$ U= \vx +  \r$$
 of a vector $\vx $  and a 2-form $ \r$
 which can be thought of as an extended
vector with $10$ components $U^I$ ($I=1,...,10$)
\begin{equation}
U^I=\left(\begin{array}{cc}
\vx ^i   \\
\r_{ij}
\end{array}\right),
\label{veca}
 \end{equation}
 where $i,j=1,..,4$ and $\r_{ij} =-\r_{ji}$.

 There is an action of $SL(5,\R)$ on $T\oplus \L^2T^* $, as follows. First, there is the natural action of
 $SL(4,\R)$ acting separately on the vector $\vx $  and  2-form $ \r$.
 There is an action of a 3-form $\Theta\in \Lambda^3T^*$
 sending 
\begin{equation}
\label{cshift}
\vx +\r \mapsto \vx +\r+\i_\vx \Theta
\end{equation}
and the action of a tri-vector $\beta\in \Lambda^3T$ with components $\b^{ijk}$
 sending 
$$\vx +\r\mapsto
\vx +\r+\i_\r \beta$$
(with $(\i_\r \beta)^i= \ha  \r _{jk}\beta^{jki}$).
These are natural generalisations of (\ref{bshift}),(\ref{beshift}).
Finally, the group closes on a scaling under which
\begin{equation}
\label{}
\vx +\r\mapsto \a ^3 \vx  + \a ^ 2 \r
\end{equation}
with $\a\in \R, \a \ne 0$.
The adjoint of $SL(5,\R)$ decomposes as 
$$24= 15+1 +4+4'$$
 under $SL(4,\R)$, corresponding to these four classes of transformation. 
 The fibres then transform in the 10-dimensional representation of $SL(5,\R)$ labelled by the index $I=1,...,10$.

An $SL(5,\R)$ bundle $\ce$ can be reduced to an $SO(5)$ bundle $\bar \ce$, and the reduction 
is equivalent to choosing an element $\cv$ of the coset $SL(5,\R)/SO(5)$, or equivalently a positive definite fibre metric $\cm$, for each point $x\in  {\cal M}$.
This can be represented by a matrix function $\cv ^A{}_I(x)$ on some patch $U\subset {\cal M}$ where $A=1,..,10$ labels the 10-dimensional representation of $SO(5)$.
Given a metric $G_{ij}$ and orientation on $\CM$, the tangent bundle becomes an $SO(4) $ bundle 
whose structure group is a subgroup of the $SO(5)$, and
the 10-dimensional representation decomposes as $10=4+6$ under $SO(4)\subset SO(5)$.

The coset
$SL(5,\R)/SO(5)$ is 14-dimensional and
 can be parameterised by a symmetric matrix
$G_{ij}$  transforming in the {\bf 10} of  $SO(4)\subset SO(5)$ and a 3-form $C_{ijk}$ transforming as a {\bf 4} of  $SO(4)$.
At each point $x\in {\cal M}$, the 
vielbein $\cv(x)$   transforms as
\begin{equation}
\cv (x) \to k(x) \cv(x) g \label{abc}  \end{equation}
under a local  $ SO(5)$ transformation $k(x)$ and rigid transformation $g\in SL(5,\R)$.
It is useful to introduce a  frame field $e^a{}_i$ for $T{\cal M}$,
so that $G_{ij}=\d _{ab} e^a{}_i e^b{}_j$ with tangent space indices $a,b...$ transforming under $SO(4)$, 
and the vielbein $e^a{}_i$ is used to convert indices $i,j...$ to $a,b...$, so that e.g. $v^a= e^a{}_iv^i$.
The local $ SO(5)$ symmetry
can be used to choose a triangular gauge for $\cv$ over some neighbourhood of ${\cal M}$, so that
\begin{equation}
\cv= \left(\begin{array}{cc} e^a{}_i & 0 \\  -e_a^{j}e_b{}^{k}  C_{ijk}
&e_a^{[i}e _b{}^{j]} \end{array}\right)
 \label{vist}  \end{equation} 
It maps $U$ given by (\ref{veca})
to 
\begin{equation}
\label{}
\bar U^A =\left(\begin{array}{cc}
u^a   \\
u_{ab}
\end{array}\right)
=\cv^A{}_I  U^I =\left(\begin{array}{cc}
v^a   \\
\r_{ab} - C_{abc}v^c
\end{array}\right)
\end{equation}

An $SO(5)$-invariant metric on sections of $\bar \ce $ is given by
\begin{equation}
\label{}
\bar \cm (\bar U , \bar U) =\bar \cm _{AB}\bar U^A \bar U^B= \d _{ab} u^a u^b + \frac 12 \d ^{ab}\d^{cd}u_{ac}u_{bd}
\end{equation}
Then
a positive definite generalised metric  $\cm$  on $\ce$ 
can be defined by (\ref{mett}) giving  the norm of (\ref{veca}) as 
\begin{equation}
\label{}
  \cm (U,U)= G(\vx ,\vx ) + G^* (\r-\i _\vx  C, \r-\i _\vx  C) \end{equation}
where $G^*$ is the norm on 2-forms constructed from $G=e^te$.  
In terms of components, this is
\begin{equation}
\label{}
  \cm (U,U) = G_{ij} \vx ^i \vx ^j + \frac 12 G^{ik}G^{jl} (\r _{ij}- C_{ijm} \vx ^m) (\r _{kl}- C_{kln} \vx ^n)
\end{equation}
so that the metric is represented by the matrix 
$\cm= \cv ^t \bar \cm \cv $ which has the form
\begin{equation}
\cm
=\left(
\begin{array}{cc}
G+ \frac 12 CG^{-1}G^{-1}C &-\frac 12 CG^{-1}G^{-1} \\
-\frac 12G^{-1}G^{-1}C   & \frac 12G^{-1}G^{-1}
\end{array}
\right)
 \label{abc}  \end{equation}
The action of the 3-form transformation on $\cv$ and $\cm$ gives
\begin{equation}
\label{}
C\mapsto C+ \Theta
\end{equation}
so that the three-form transformation shifts the three-form field $C$.

\subsection{$n=5$, $E_5=Spin(5,5)$}

Consider next the case of a five-manifold, with $E_5=Spin(5,5)$. 
In this case, in addition to the 2-form, a 5-form is added to the fibres.
The bundle  $T\oplus T^*$ is then replaced with $  T\oplus \L^2T^* \oplus \L ^5 T^*$ 
with 16-dimensional fibres 
transforming in the $5+10'+1$ representation of $SL(5,\R)$.
 A section is then a formal sum
 $$ U= \vx +  \r+\s$$
 of a vector $\vx $, a 2-form $ \r$   and  a 5-form $\s$.
 Given a volume form $\e \in \L^5T^*$ and its dual $\ti \e \in \L^5T$ with $\i_{\ti \e} \e=1$, 
 this is equivalent to
  to the sum of a 0-form $*\s= \i_{\ti \e} \s
  $, a 2-form $\r$ and a 4-form $*\vx= \i_v \e $, and so  there is a natural action of
 $Spin(5,5)$ on this under which the fibres transform as ${\bf 16 ^+}$, the positive chirality spinor representation.
 The adjoint of $Spin(5,5)$ decomposes under $SL(5,\R)$ as
 $${\bf 45 = 24+1 +10+10'}$$
 consisting of the natural action of $SL(5,\R)$ on   tangent vectors and forms on a 5-fold, 
a scaling transformation and the action of a  3-form $\Theta _{ijk}$ and a 3-vector $\b^{ijk}$, so that this is very similar to the $n=4$ case.
The coset space   $Spin(5,5)/ H_5$ where $H_5=(Spin(5)\times Spin(5))/\Z_2$ has dimension 25
and can be parameterised by a symmetric matrix  $G_{ij}$ and 3-form $C_{ijk}$.
Then as for $n=4$, there is a generalised metric $\cm(x)$ and vielbein $\cv$ parameterised by a metric
$G_{ij}(x)$ and 3-form $C_{ijk}(x)$ on ${\cal M}$, with the 3-form transforming as $C\mapsto C+ \Q$.

 \subsection{$n=6$, $E_6$}
 
 As for $n=5$, the bundle  $T\oplus T^*$ is   replaced with $  T\oplus \L^2T^* \oplus \L ^5 T^*$ 
with 27-dimensional fibres 
transforming in the $6+15+6$ representation of $SL(6,\R)$, with a natural action of $E_6$ acting in the {\bf 27} representation.
A section is   a 27-dimensional vector  decomposing as a formal sum
 $$ U= \vx +  \r+\s$$
 of a vector $\vx $, a 2-form $ \r$   and  a 5-form $\s$.
The  adjoint of $E_6$ decomposes under $SL(6,\R)$ as
 $${\bf 78 = 35+1 +20+20+1+1}$$
  consisting of the natural action of $SL(5,\R)$ on   tangent vectors and forms on a 5-fold, 
a scaling transformation, the action of a  3-form $\Theta _{ijk}$ and a 3-vector $\b^{ijk}$, as before,
but now in addition there is     the action of a 6-form $\ti \Theta \in \L^6T^*$ and a 6-vector $\ti \b \in \L^6T$; these are singlets, but 
  regarding them as 6-forms and 6-vectors is suggested by the fact that 6-forms and 6-vectors arise for $n=7$.
The coset $E_6/H_6$ where $H_6=Sp(4)/\Z_2$ is 42-dimensional and can be parameterised by 
a symmetric matrix  $G_{ij}$, a  3-form $C_{ijk}$  and a 6-form $\ti C_{i_1...i_6}$ (dual to a scalar in 6 dimensions). Then   the generalised metric $\cm(x)$ and vielbein $\cv$ are parameterised by a metric
$G_{ij}(x)$, 3-form $C_{ijk}(x)$ and a 6-form $\ti C_{i_1...i_6}(x)$ on ${\cal M}$. 
 The group $E_6$ has a maximal subgroup $SL(6,\R)\times SL(2,\R)$ under which
 $${\bf 27} \to {\bf (6,2)} + {\bf (15,1)}, \qquad
{\bf 78} \to {\bf (35,1)} + {\bf (20,2)}+ {\bf (1,3)}$$

 For $n=7$, as will be seen below,   
 the bundle  $T\oplus T^*$ is   replaced with $  T\oplus \L^2T^* \oplus \L ^5 T^*\oplus \L ^6 T$,
 suggesting that for $n=6$ one also consider a generalisation in which $\L ^6 T$ is added to the 
 generalised tangent bundle. Then $\L ^6 T$ is invariant under $E_6$ and $SL(6,\R)$, so that $E_6$ acts on  
  $  T\oplus \L^2T^* \oplus \L ^5 T^*\oplus \L ^6 T$ in the ${\bf 27+1}$ representation.
 This extra singlet corresponds to an extra  charge that is allowed by the supersymmetry algebra \cite{Hull:2000cf}. It is
 not known whether states carrying this charge arise in M-theory, but if they do, their presence would have dramatic implications \cite{Hull:2000zn}.
 
  \subsection{
$n=7$, $E_7$} \label{n=7}

 For $n=7$, the bundle  $T\oplus T^*$ is   replaced with  
 $$  T\oplus \L^2T^* \oplus \L ^5 T^*\oplus \L ^6 T$$
  with 56--dimensional fibres 
transforming in the $7+21'+21+7'$ representation of $SL(7,\R)$, with a natural action of $E_7$ acting in the {\bf 56} representation.
$E_7$ has a maximal $SL(8,\R)$ subgroup, and these $SL(7,\R)$ representations combine into the $28+28'$ of $SL(8,\R)$.
  A section is   a 56-dimensional vector  decomposing as a formal sum
 $$ U= \vx +  \r+\s+\t$$
 of a vector $\vx $, a 2-form $ \r$,  a 5-form $\s$   and  a 6-vector $\t$.
 
 The adjoint of $E_7$ decomposes under $SL(7,\R)$ as 
 $${\bf 133 = 48+1+ 35+35'+7+7'}$$
 and so in addition to the standard action of $SL(7,\R)$ and a scaling, there is the action of a 3-form $\Theta\in \L^3T^*$, a 3-vector $\b \in \L^3T$, a 6-form $\ti \Theta\in \L^6T^*$ and  a 6-vector $\ti \b \in \L^6T$. The action of the  6-form and 6-vector combine with the    action of $SL(7,\R)$ and the scaling
 to generate an $SL(8,\R)$ subgroup.
 The coset $E_7/H_7$ where $H_7=SU(8)/\Z_2$ is 70-dimensional and can be parameterised by 
a symmetric matrix  $G_{ij}$, a  3-form $C_{ijk}$  and a 6-form $\ti C_{i_1...i_6}$
($70=28+35+7$).
Then   the generalised metric $\cm(x)$ and vielbein $\cv$ is specified in terms of a metric
$G_{ij}(x)$, 3-form $C_{ijk}(x)$ and a 6-form $\ti C_{i_1...i_6}(x)$ on ${\cal M}$.

The action of $E_7$ can be understood as follows.
Consider the 8-manifold $N={\cal M}\times S^1$ with the natural $U(1)$ action generated by a vector $k$ tangent to $S^1$; let $\ti k$ be the dual one-form on $S^1$, with $\ti k (k)=1$. If $\th \sim \th + 2\p$ is the $S^1$ coordinate, then
$k=\pa/\pa \th$ and $\ti k = d \th$.
A 2-form $\f$ on $N$ is specified by  a 1-form $\f_1= \i
_k \f$ and a 2-form $\f_2=\f- \ti k\we \i_k\f$ with $\i_k\f_2=0$, and if $\f$ is $U(1)$ invariant (i.e. the Lie derivative ${\cal L}_k\f=0$) these pull-back to a 1-form  $\f_1'$ and  2-form $\f_2'$ on ${\cal M}$. 
We can then define a 2-form and 6-vector on ${\cal M}$ by
$\r =\f_2'$ and $\t = * \f_1'= \i _{ \f_1'} \ti \e$ where 
$\ti \e$ is the 7-vector on  ${\cal M}$ dual to the volume form $\e$.
The natural action of $SL(8,\R)$ on $\L^2 T^*N$ 
gives the action of $SL(8,\R)$ on $\f$  and hence on 
$\r,\t$ which transform according to the ${\bf{28'}}$ representation.
Similarly, an invariant bi-vector $\chi  \in \L^2 TN $ gives a vector $\chi  _1' \in T{\cal M}$  and 
a bi-vector $\chi  _2' \in \L^2T{\cal M}$, and   these define a vector 
 $\vx=\chi  _1'$ and a 5-form $\s= *\chi   _2' = \i _{\chi  _2'} \e$. The action of $SL(8,\R)$ on $\L^2 TN$ 
 then gives the $SL(8,\R)$ transformations of
 $\vx , \s$ which combine into the {\bf{28}} representation.
 
 The remaining generators of $E_7$ combine into a   4-form on $N$,  $\SS \in \L^4 T^*N$.
 The infinitesimal action $U(\SS) $ of $\SS$  on $\f+\chi   \in \L^2 T^*N \oplus \L^2 TN$ 
 is 
 \begin{equation}
\label{sgsfg}
U(\SS): \f+\chi   \mapsto \f+\chi   + \i _\chi   \SS + \i_\f *\SS
\end{equation}
where $*\SS$ is the dual on $N$,     $*\SS= \i _\SS (k\wedge \ti \e) \in \L^4 TN$.
The 4-form $\SS$ gives a 3-form  $\Q$ and 4-form $\b '$ on ${\cal M}$, and the
4-form $\b '$ dualises to a 3-vector $\b \in \L^3 T{\cal M}$  (given by $\b = *\b' =
\i_{\b'} \ti \e$).
Then the transformation (\ref{sgsfg})
 gives the infinitesimal  transformation of $ U= \vx +  \r+\s+\t$ under the
 action of the 3-form $\Q$ and 3-vector $\b$.
 The corresponding transformations under $E_n$ in dimensions $n< 7$ follow by truncation.
 
The vielbein $\cv$ is constructed following \cite{Cremmer:1979up}, and can be parameterised in terms of $G_{ij}$, the 3-form $C_{ijk}$ and a vector
  $B^j= \frac 1 {6!} \e ^ {ji_1...i_6} \ti C_{i_1...i_6}$
in the factorised form
\begin{equation}
\cv = \a\b  \exp [ U(C\we k)]
\end{equation}
Here   $U(C\we k)$ is the map (\ref{sgsfg}) with 
$$\SS=C\we k$$
This $\SS$ has components $\SS_{IJKL}$ where $I=1,...,8$ label coordinates on $N$ which satisfy 
\begin{equation}
(*\SS)^{IJKL} \SS_{KLMN}(*\SS)^{MNPQ}=0
\end{equation}
and as a result $U(C\we k)$  is nilpotent,
$$[U(C\we k)]^4=0$$
Then the exponential becomes the polynomial
\begin{equation}
\exp [ U(C\we k)] = 1+U+ \frac 12 U^2 +  \frac 16 U^3
\end{equation}
and so $\cv$ is cubic in the 3-form $C$.
The $\a.\b$ are $SL(8,\R)$ transformations acting in the $28+28'$ representation.
Their action in the fundamental 8-dimensional representation
are given by $8\times 8$ matrices $\a^I{}_J,\b^I{}_J$ which take the $(7+1)\times (7+1)$  block form
\begin{equation}
\a= \left(\begin{array}{cc}
e^a{}_j &  0\\
  0  & e^{-1}
\end{array}\right),
\qquad
\b= \left(\begin{array}{cc}
\d^i{}_j & B ^i \\
  0  & 1
\end{array}\right),
\label{dsfsdh}
 \end{equation}
where $e^a{}_i$ is a vielbein for $M$ with
$e^a{}_i e^b{}_j \d_{ab}= G_{ij}$ and $e = \det (e_i{}^a) = \sqrt {\det (G_{ij})}$.
The generalised metric is then given by
$$\ch= \cv^t \cv$$
 and is polynomial in both  $C$ and $\ti C= *B$.
 
\section{Type M Extended Tangent Bundles and Extended Spin Bundles}

 The bundles identified in the last section are summarised in table 2, with a natural action of $E_n$ on the fibres in the representation {\bf{R}}. The coset $E_n/H_n$ is parameterised by the fields given in table 3.

  \begin{table}
\begin{center}
\begin{tabular}{|c|cccc|}
\hline
n & $E_{n}$ & ${\bf R}$ & $SL(n,\R)$ reps & $\ce$ \\
\hline
2  & $SL(2,\R) \times \R$ &   2+1& 2+1& $\ce\sim T\oplus \L^2T^*  $\\
3  & $SL(3,\R) \times SL(2,\R)$ & $(3,2) $&$ 3+3 $&$\ce\sim T\oplus \L^2T^*  $ \\
4  & $SL(5,\R)$ & $10 $ & $ 4+6' $& $\ce\sim T\oplus \L^2T^*   $ \\
5  & $Spin(5,5)$ & $16 $ &  $ 5+10'+1 $ & $\ce\sim T\oplus \L^2T^*  \oplus \L^5T^* $\\
6  & $E_{6(6)}$ & $27  (+1)$& $ 6+15'+6 (+1)$& $\ce\sim T\oplus \L^2T^*  \oplus \L^5T^*( \oplus \L^6T )$ \\
7  & $E_{7(7)}$ &$56  $ & $ 7+21'+21+7' $ & $ \ce\sim T\oplus \L^2T^*  \oplus \L^5T^* \oplus \L^6T $ \\
\hline    
\end{tabular}
\caption{The bundle $\ce$ over an $n$-dimensional manifold ${\cal M}$ has fibre in the representation ${\bf R}$ of $E_{d+1}$. The decomposition into $SL(n,\R)$ representations gives a corresponding decomposition of $\ce$.
} \label{tab2} 
\end{center}
\end{table}

 \begin{table}
\begin{center}
\begin{tabular}{|c|cccc|}
\hline
n & $E_n$ &  dim($E_n$)& dim($E_n/H_n$) & Coset moduli\\
\hline
2  & $SL(2,\R) \times \R$ &  4& 3 & $G$\\
3  & $SL(3,\R) \times SL(2,\R)$ &   11&7 =6+1& $G,C_3$\\
4  & $SL(5,\R)$ & 24& 14 =10+4& $G,C_3$\\
5  & $Spin(5,5)$ &   45& 25 =15+10& $G,C_3$\\
6  & $E_{6(6)}$ &   78&42=21+20+1 & $G,C_3,C_6$\\
7  & $E_{7(7)}$ &  133 & 70=28+35+7 & $G,C_3,C_6$\\

\hline    
\end{tabular}
\caption{The U-duality groups $E_n$,   the     dimensions of the cosets $E_n/H_n$ and the parameterisation of the cosets in terms of a metric $G$, a 3-form $C_3$ and a 6-form $C_6$.} \label{tab3} 
\end{center}
\end{table}

 As discussed in section \ref{Geometry}, $T\oplus T^*$ strictly speaking has structure group $GL(d,\R)$, and this can be 
 extended by twisting with a gerbe  to a generalised tangent bundle with structure group
 $GL(d,\R)\ltimes \W^{2,cl}$, where 
 $\W^{2,cl}$ is  the bundle of closed 2-forms, and this  preserves the Courant bracket.
 In section
 \ref{gengen}, this was generalised further to type I extended tangent bundles with structure group $O(d,d)$. This will not preserve 
 the Courant bracket in general, but such   structures are relevant for non-geometric backgrounds in string theory.
 
 In the same way, the bundle
  $$  T\oplus \L^2T^* \oplus \L ^5 T^*\oplus \L ^6 T$$
has structure group $GL(d,\R)$.
This can again be twisted with a gerbe, in a similar way to section \ref{gerbe}.

Consider first $T\oplus \L^2 T^*$.
The 3-form $C$ can be taken to be a connection with transition functions
$$(\d C)_{\a\b}\equiv C_\beta-C_\alpha=d\l_{\alpha\beta}$$ 
for some 2-forms $\l_{\alpha\beta}$ on the overlaps $U_{\alpha}\cap U_{\beta}$
with
consistency conditions
$$(\d \l)_{\alpha\beta\d}\equiv \l_{\alpha\beta}+\l_{\beta\gamma}+\l_{\gamma\alpha}= d\k_{\alpha\beta\gamma}$$
for 1-forms $ \k_{\alpha\beta\gamma}$ on triple overlaps $U_{\alpha}\cap U_{\beta}\cap U_{\gamma}$. These satisfy
\begin{equation}
\label{}
(\d \k)_{\alpha\beta\gamma\d} \equiv
\k_{\alpha\beta\gamma} +\k_{\beta\gamma\d}+\k_{\gamma\d\a}+\k_{\d\alpha\beta}=g_{\a\b\g\d}^{-1} dg_{\a\b\g\d}
\end{equation}
for some maps  $g_{\a\b\g\d}:U_\alpha\cap U_\beta\cap U_\gamma \cap U_\delta \to U(1)$ from quadruple overlaps to $U(1)$, which in turn  satisfy
\begin{equation}
\label{}
g_{\a\b\g\d}g_{\b\g\d\e}g_{\g\d\e\a}g_{\d\e\a\b}g_{ \e\a\b\g}=\1
\end{equation}
on quintuple overlaps.
 (This could be generalised to allow   
   \begin{equation}
\label{}
\k_{\alpha\beta\gamma} +\k_{\beta\gamma\d}+\k_{\gamma\d\a}+\k_{\d\alpha\beta}= d\f_{\a\b\g\d}
\end{equation}
for some 0-forms $\f_{\a\b\g\d}$ on quadruple overlaps satisfying a consistency condition $(\d \f)_{\a\b\g\d\e\eta}=c_{\a\b\g\d\e\eta}$
on quintuple overlaps for constants $c_{\a\b\g\d\e\eta}$ satisfying $(\d c)_{\a\b\g\d\e\eta\k}=0$.)

The    $\l_{\alpha\beta}$ can be used to define a bundle $E$ over $M$
by identifying $T\oplus \L^2 T^*$ on $U_{\alpha}$ with $T\oplus \L^2 T^*$ on $U_{\beta}$ by the C-field action $\vx+\r \mapsto \vx +\r+i_vd\l_{\alpha\beta}$.   The fibre over a point $x$ in $M$ is again
$T_x\oplus \L^2 T^*_x$, but   the transition functions are now in
 $GL(d,\R)\ltimes \W^{3,cl}$, where 
 $\W^{3,cl}$ is  the bundle of closed 3-forms. This preserves the Courant bracket on  $T\oplus \L^2 T^*$.

This extends to  $  T\oplus \L^2T^* \oplus \L ^5 T^*\oplus \L ^6 T$, with
the 6-form $\ti C$ a connection with transition functions
$$(\d \ti C)_{\a\b}\equiv \ti C_\beta-\ti C_\alpha=d\ti \l_{\alpha\beta}$$ 
for some 5-forms $\ti \l_{\alpha\beta}$  on the overlaps $U_{\alpha}\cap U_{\beta}$ satisfying similar consistency conditions to the above.
The group $E_n$ has a subgroup containing $GL(n,\R)$ and transformations generated by a 3-form and (for $n=6,7$) a 6-form, and the fibres 
 $$ U= \vx +  \r+\s+\t$$
can be patched together on overlaps using such transformations with closed 3-form and 6-form
generators. The structure group is then generated by
$GL(d,\R)$, $ \W^{3,cl}$ and $ \W^{6,cl})$, where 
 $\W^{p,cl}$ is  the bundle of closed $p$-forms.
 The action of the 3-forms and 6-forms generates a non-trivial algebra; if $\d_3(\L)$ is the transformation generated by a closed 3-form $\L$ and $\d_6(\SS)$ is the transformation generated by a closed 6-form $\SS$, then these satisfy an algebra $\ca$
 whose only non-trivial commutation relation is \cite{Cremmer:1998px}
 \begin{equation}
[\d_3(\L),\d_3 (\L ')]= \d_6(\L \we \L ')]
\end{equation}
Then the structure group is $GL(d,\R) \ltimes \ca$.

 To incorporate non-geometric backgrounds, the bundle
 $  T\oplus \L^2T^* \oplus \L ^5 T^*\oplus \L ^6 T$  with transition functions in $GL(n,\R)$
 (or its generalisation twisted by  gerbes with structure group 
$GL(d,\R)\ltimes ( \W^{3,cl}\oplus \W^{6,cl})$)
  is generalised to 
 a vector bundle $\ce$  over the $n$-dimensional oriented manifold $\cal M$ with 
 structure group $E_n$  and fibres in the representation {\bf{R}}
   given in table \ref{tab2} for each value of $n$.
  This will be referred to as  an extended tangent  
  bundle. 
  In general, the transition functions will mix the metric $G$ with the gauge fields $C_3,C_6$, so that these 
  will be defined locally in patches 
through the choice of $\cv _\a$, but will not patch together to form 
     tensor fields or gerbe connections.

As discussed in section \ref{GenEx}, the extended tangent bundle  $\ce$ with structure group $E_n$ can be reduced to      a
     bundle $\bar \ce$ with structure group $H_n$, the maximal compact subgroup of $E_n$ given in table 1, and the reduction is equivalent to a choice of vielbein $\cv$.
    The groups $H_n$ each have   a natural double cover $\ti H_n$
    given in table 4. The various $\Z _2$ factors and double cover maps are given in \cite{Keurentjes:2003hc}.
    An M-type extended spin bundle  $\ti \ce$ is a bundle over ${\cal M}$ that projects onto $\bar \ce$ under the projection
 $p:\ti H_n\to H_n$, and a necessary and sufficient condition for this is that $w_2(\bar \ce) =0$.

      \begin{table}
\begin{center}
\begin{tabular}{|c|ccc|}
\hline
n & $E_n$ & $H_n$ &  $\ti H_n$  \\
\hline
2  & $SL(2,\R) \times \R$ & $SO(2)$ & $Spin(2)$\\
3  & $SL(3,\R) \times SL(2,\R)$ & $SO(3) \times SO(2)$& $Spin(3) \times Spin(2)$ \\
4  & $SL(5,\R)$ & $SO(5)$ & $Spin(5)$\\
5  & $Spin(5,5)$ & $(Sp(2) \times Sp(2))/\Z_2$&   $Sp(2) \times Sp(2)$\\
6  & $E_{6(6)}$ & $Sp(4)/\Z_2$ &  $Sp(4)$ \\
7  & $E_{7(7)}$ & $SU(8)/\Z_2$ & $SU(8)$\\
8  & $E_{8(8)}$ & $Spin(16)/\Z_2$ & $Spin(16)$\\
\hline    
\end{tabular}
\caption{The U-duality groups $E_n$, their maximal compact
  subgroups $H_n$,  and the   double covers $\ti H_n$ of $H_n$.}
   \label{tab4} 
\end{center}
\end{table}

\section{Type II Geometries}\label{type2}

In this section, the generalisations of  generalised geometry suggested by type II string theory
on a $d$-dimensional manifold $M$ are studied, in which  
$T\oplus T^*$ with a natural action of $SO(d,d)$ is replaced by
$\ce ^\pm \sim T\oplus T^*\oplus S^\pm  \oplus \dots $ with a natural action of $E_{d+1}$.
The positive chirality spin bundle $S^+$ is used for   type IIA string backgrounds and 
the negative chirality spin bundle $S^-$ is used for   type IIB string backgrounds,
so $\ce^+$ will be referred to as a type IIA geometry and
$\ce^-$ will be referred to as a type IIB   geometry.
For a given embedding $SO(d,d)\subset E_{d+1}$, the two choices of chirality give 
two distinct representations ${\bf R}^\pm$ of $E_{d+1}$.
Equivalently, one could fix the representation ${\bf R}$ of   $E_{d+1}$ and choose two different embeddings
$SO(d,d)\subset E_{d+1}$ to obtain   two decompositions $\ce   \sim T\oplus T^*\oplus S^\pm  \oplus \dots $.
 
\subsection{$d=3$, $E_4=SL(5,\R)$}

Consider first the case $d=3$, with $E_4=SL(5,\R)$. 
For $\ce ^+$, we take the fibres to be in the {\bf 10} representation.
 Under the $SL(4,\R)= Spin(3,3)$ subgroup, 
 the {\bf 10} of $SL(5,\R)$ decomposes as
 $ {\bf 10} \to  {\bf 6}+  {\bf 4}$, corresponding to the vector   and negative chirality spinor  representations
 of $ Spin(3,3)$. Under $SL(3,\R)\subset Spin(3,3)$, the  {\bf 6} decomposes into the 
 $ {\bf 3}+  {\bf 3'}$, and the  {\bf 4} decomposes into the 
 $ {\bf 1}+  {\bf 3}$.
 Then  locally the fibres of $\ce^+$ decompose into $T\oplus T^* \oplus \L^0 T^* \oplus  \L^2 T^*$. A section is then a formal sum
 $$ U= \vx +\x + \r_0 +\r_2$$
 of a vector $\vx $, a 1-form $\x$, a 0-form $\r_0$ and a 2-form $ \r_2$.
 The 0-form $\r_0$ and  2-form $ \r_2$ combine to form a positive chirality spinor
 of $Spin(3,3)$, so that $U$ is the sum of a vector $V=\vx +\x $ and a spinor $\r^+=\r_0+ \r_2$
 of $Spin(3,3)$.
 
 Similarly, for $\ce ^-$, we take the fibres to be in the dual ${\bf 10'}$ representation, decomposing as
  $ {\bf 10'} \to  {\bf 6}+  {\bf 4'}$ under $SL(4,\R)= Spin(3,3)$, and further decomposing into $SL(3,\R)$ representations gives
    $ {\bf 3}+  {\bf 3'}+{\bf 1}+  {\bf 3'}$. Then a section is a formal sum  
 $$ U= \vx +\x + \r_1 +\r_3$$
 of a vector $v$, a 1-form $\x$, a 1-form $\r_1$ and a 3-form $ \r_3$, 
 in 
 $$T\oplus T^* \oplus T^* \oplus  \L^3 T^* \sim T\oplus T^* \oplus S^-
 $$
 The 1-form $\r_1$ and  3-form $ \r_3$ combine to form a negative chirality spinor
 of $Spin(3,3)$, so that $U$ is the sum of a vector $V=\vx +\x $ and a spinor $\r^-=\r_1+ \r_3$
 of $Spin(3,3)$. 
 
 The adjoint of $SL(5,\R)$ decomposes under $Spin(3,3)$ as
 \begin{equation}
\label{}
24= 15+1 +4^++4^-
\end{equation}
with two spinor generators $\Theta^\pm \in S^\pm$.
In addition to the standard action of $Spin(3,3)$  and a scaling transformation, there are two extra generators in spin representations of  $Spin(3,3)$ that transform $T\oplus T^*$ and $S^\pm$ into one another.
The coset space $SL(5,\R)/SO(5)$ is 14-dimensional and can be parameterised by
a metric $G_{ij}$, 2-form $B_{ij}$ and scalar $\Phi$, together with either
even forms $C_0,C_2$ combining into a positive chirality spinor $C^+$, or
odd forms $C_1,C_3$ combining into a positive chirality spinor $C^-$. These two possibilities correspond to two gauge choices for the local $SO(5) $.
The parametrisation in terms of $C^+$ is useful for the IIA string and that in terms of $C^-$  for the IIB string.
The generators $\Q^\pm$ act as shifts of $C^\pm$, $$C^\pm \mapsto C^\pm+\Q^\pm$$
 
 \subsection{General $d\le 6$}
 
 A similar structure applies for other $d\le6$, as summarised in table 3.
  \begin{table}
\begin{center}
\begin{tabular}{|c|cccc|}
\hline
d & $E_{d+1}$ & ${\bf R}$ & $Spin(d,d)$ reps & $\ce$ \\
\hline
2  & $SL(3,\R) \times SL(2,\R)$ & $(3,2) $&$ 4+ 2^\pm $&$\ce\sim T\oplus T^*\oplus S^\pm $ \\
3  & $SL(5,\R)$ & $10$ & $ 6+ 4^\pm $& $\ce\sim T\oplus T^*\oplus S^\pm $ \\
4  & $Spin(5,5)$ & $16$ &  $ 8+ 8^\pm $ & $\ce\sim T\oplus T^*\oplus S^\pm $ \\
5  & $E_{6(6)}$ & $27 (+1)$& $ 10+1 (+1)+ 4^\pm $ & $\ce\sim T\oplus T^* \oplus \L^5 T^*(\oplus \L^5 T)\oplus S^\pm $  \\
6  & $E_{7(7)}$ &$56 $ & $ 12+12+ 32^\pm $ & $\ce\sim T\oplus T^*  \oplus \L^5 T\oplus \L^5 T^* \oplus S^\pm$\\
\hline    
\end{tabular}
\caption{The bundle $\ce$ over a $d$-dimensional manifold $M$ has fibre in the representation ${\bf R}$ of $E_{d+1}$. The decomposition into $Spin(d,d)$ representations gives a corresponding decomposition of $\ce$. The upper sign is for the IIA geometry and the lower one for IIB geometry.
} \label{tab5} 
\end{center}
\end{table}
 For $d=2,3,4$, $T\oplus T^*$   is extended to  
$$\ce ^\pm = T\oplus T^*\oplus S^\pm   $$ with a natural action of $E_{d+1}$.
For example, for $d=4$,  the fibre is in the positive chirality spinor representation $16^+$ of $E_5=Spin(5,5)$ for the IIA geometry. Under the natural
 embedding of $Spin (4,4)\subset Spin (5,5)$, the $16^+$  decomposes into 
 the spinor representations $8^++8^-$ of $Spin (4,4)$.
 This is related by $Spin (4,4)$ triality to an embedding in which
 it decomposes into a vector and spinor $8_v+8^+$, and this is the embedding 
 used here, with $\ce ^+\sim T\oplus T^*\oplus S^+ $. For type IIB,   $\ce\sim T\oplus T^*\oplus S^- $, and this can either be regarded as coming from the same embedding of $Spin (4,4)\subset Spin (5,5)$ but with the fibres in the negative chirality spinor representation $16^-$ of $Spin(5,5)$, or  as arising from 
 keeping the same $16^+$  representation but choosing a different embedding of $Spin (4,4)\subset Spin (5,5)$ (related by triality to the other two embeddings discussed above).
 
 For $d=5,6$, $T\oplus T^*$   is extended to  
 $$\ce\sim T\oplus T^*  \oplus \L^5 T\oplus \L^5 T^* \oplus S^\pm$$ transforming under $E_d$,
 with $\L^5 T$ corresponding to NS 5-brane charge and  $\L^5 T^*$ corresponding to KK monopole charge.
 For $d=5$, this is the reducible $27+1$ representation, and the $ \L^5 T$ factor can be removed to leave the $27$.
 For $d=6$, $E_7$ has a maximal subgroup $SO(6,6)\times SL(2,\R)$, and under this the 
 $56$ decomposes as
 $56=(12,2)+ (32,1)$.
 As $ \L^5 T\oplus \L^5 T^* \sim T^*\oplus T  $ for $d=6$, 
 $$\ce\sim T\oplus T^*  \oplus  T\oplus  T^* \oplus S^\pm$$
 and $T\oplus T^* $ forms an $SL(2,\R)$ doublet with $\L^5 T\oplus \L^5 T^*$, with both in the 12-dimensional vector representation of $SO(6,6)$.

The decomposition of the adjoint of $E_{d+1}$ into $Spin (d,d)$ representations is given in  table 5. In each case there are two spinor generators, which are of the same chirality for $d$ even and opposite chiralities for odd $d$.
Convenient  parameterisations 
 of the coset space $E_{d+1}/H_{d+1}$ are also given.
 For each $d$, these are represented by
 $$ G,B,\ti B, \Phi, C^\mp$$
  including the $d^2$ parameters assembled into the metric $G$ and 2-form $B$, a scalar $\Phi $, and a 6-form $\ti B$ which only contributes for $d=6$, corresponding to a 6-form field dual to the 2-form $B$.
   In addition, for the IIA theory there is a negative chirality spinor $C^-$ corresponding to a set of odd forms
   $C^-\sim C_1,C_3,C_5$, while for the IIB theory there is a positive chirality spinor $C^-$ corresponding to a set of even forms
   $C^+\sim C_0,C_2,C_4,C_6$.

   \begin{table}
\begin{center}
\begin{tabular}{|c|cccc|}
\hline
d & $E_{d+1}$ & Adjoint  & $Spin(d,d)$ decomposition  & $E_{d+1}/H_{d+1}$ parameterisation \\
\hline
2  & $SL(3,\R) \times SL(2,\R)$ & $8+3$&$ 6+ 1+2^\mp +2^\mp $&$7=6+1+2^\mp $ \\
3  & $SL(5,\R)$ & $24$ & $ 15+1+ 4^++4^- $& $14=9+1+4^\mp $ \\
4  & $Spin(5,5)$ & $45$ &  $ 28+1+ 8^\mp+8^\mp $ & $25=16+1+8^\mp  $ \\
5  & $E_{6(6)}$ & $78$& $ 45+1+ 16^++16^- $ & $42=25+1+16^\mp $  \\
6  & $E_{7(7)}$ &$133$ & $ 66+1+1+1+ 32^\mp+32^\mp$ & $70=36+1+1+32^\mp $\\
\hline    
\end{tabular}
\caption{The bundle $\ce$ over a $d$-dimensional manifold $M$ has fibre in the representation ${\bf R}$ of $E_{d+1}$. The decomposition into $Spin(d,d)$ representations gives a corresponding decomposition of $\ce$.
} \label{tab6} 
\end{center}
\end{table}

 \subsection{Reduction of M-Geometries to Type IIA Geometries}

Consider an M-geometry on an $n$-dimensional manifold $\CM$  which is a circle bundle over a 
$d=n-1$-dimensional manifold $M$, with $n\le 7$.
As in subsection \ref{n=7}, each $p$-form on $\CM$ that is invariant under the circle action projects to a 
$p$-form and a $p-1$-form on $M$, and each invariant  $p$-vector on $\cm$  projects to a 
$p$-vector and a $p-1$-vector on $M$.
Thus
\begin{equation}
\L ^p T\CM \vert _{U(1)}  \sim \L ^p TM \oplus  \L ^{p-1} TM,
\qquad
\L ^p T^*\CM \vert _{U(1)}  \sim \L ^p T^*M \oplus  \L ^{p-1} T^*M
\label{abc}
\end{equation}
The M-geometry on $\CM$
is based on
 $$  T\oplus \L^2T^* \oplus \L ^5 T^*\oplus \L ^6 T$$
 For invariant forms and multi-vectors, this reduces to the following structure on $M$:
 \begin{equation}
T\oplus T^*  \oplus \L^5 T\oplus \L^5 T^* \oplus [ \L^0 T^* \oplus \L^ 2T^* \oplus \L^ 4T^* \oplus \L^ 6T^* ]
 \sim
 T\oplus T^*  \oplus \L^5 T\oplus \L^5 T^* \oplus S^+
\label{abc}
\end{equation}
This uses that for $d=6$, $ \L ^6 TM\sim   \L ^6 T^*M$, while for $d<6$, $ \L ^6 TM$ does not arise.

The generalised metric on $\CM$ is parameterised by a metric $G$, a 3-form $C$ and a 6-form $\ti C$. If these are invariant under the circle action, then the 3-form projects to a 2-form $B$ and 3-form $C_3$, the 6-form $\ti C$ gives  a 6-form $ \ti B$ and a 5-form $ C_5$, while the metric projects to a metric $G_M$, 1-form $C_1$ and scalar $\Phi$ on $M$.
In this way the M-geometry generalised metric $\ch (G, C,\ti C)$ on $\CM$ gives rise to the IIA-geometry generalised metric 
$\ch(G_M,B,\ti B, \Phi, C_1,C_3, C_5)$ depending on the IIA-geometry fields on a manifold of dimension $d\le 6$:
\begin{equation}
\{ G_M,B,\ti B, \Phi, C_1,C_3, C_5 \}
\sim 
\{ G_M,B,\ti B, \Phi, S^- \}
\label{abc}
\end{equation}
The explicit parameterisation of the M-geometry generalised metric $\ch$ on a 7-fold $\CM$ given in subsection \ref{n=7} then gives that of the type-IIA generalised metric on a 6-fold $M$, and the parameterisation of the type-IIA generalised metric for $d<6$ follows by truncation.

 \section{Type II Extended Tangent Bundles and Extended Spin Bundles}

 The type II geometries have extended tangent spaces of the form
 $\ce\sim T\oplus T^*   \oplus \L^\pm$
 or
 $$\ce\sim T\oplus T^*  \oplus \L^5 T\oplus \L^5 T^* \oplus \L^\pm$$
 where 
 $\L^\pm$ are the bundles of even or odd forms, and these have structure group 
 $GL(d,\R)$.
 The coset space is parameterised by the fields $ G,B,\ti B, \Phi, C^\mp$. This structure can be twisted by  gerbes by allowing  the $p$-form fields to be gauge fields with transition functions that are
 closed  $p$-forms. The action of $E_{d+1}$ includes transformations generated by $p$-forms for the same values of $p$ that act as shifts of the $p$-form gauge fields
 and so can be used in the transition functions in the same way as in section 5.
As in section 5, this can be generalised to allow 
general vector bundles with structure group $E_{d+1}$ and with fibres in the representations ${\bf{R^\pm}}
 $ given in table \ref{tab5}.
 Again, this will give a non-geometric construction in general, as the transition functions will mix the metric with the various gauge fields.
 The type II extended tangent bundle $\ce_{d+1}$ over a $d$-manifold $M$ with structure group $E_{d+1}$ reduces to a bundle $\bar \ce$ with
compact structure group $H_{d+1}$, and a type II generalised 
 spin-bundle is 
 a bundle $\ti \ce$  with
  structure group $\ti H_{d+1}$, the double cover of $H_{d+1}$ from table 4, that projects onto $\bar \ce$ under the double cover map.

\section{Special Structures, Generalised Holonomy and Supersymmetry}

In Riemannian geometry, interesting classes of geometry are characterised by specifying the holonomy
of the Levi-Civita connection. In $n$ dimensions, a general space will have 
holonomy $O(n)$, but a Kahler space has holonomy $U(n/2)$ (for $n$ even), a Calabi-Yau space has
holonomy $SU(n/2)$ (for $n$ even), and special holonomies $G_2$ and $Spin(7)$ can arise for $n=7,8$ respectively.
There is an intimate relation between the holonomy and the number of covariantly constant  spinors, and hence the number of supersymmetries preserved when the geometry is used in a 
supergravity solution.

In generalised geometry, interesting classes are given by generalised complex \cite{Hitchin:2004ut}, generalised Kahler \cite{Gualtieri} and generalised  Calabi-Yau geometries \cite{Hitchin:2004ut}, and these too are related to supersymmetry \cite{Lindstrom:2004eh} 
--\cite{Bredthauer:2006hf}.

In previous sections, extended tangent and spin bundles of types I,II and M were discussed, and geometries specified by a metric G and various antisymmetric tensor fields.
In this section, connections on the extended spin bundle that are constructed from this geometrical data will be discussed, and interesting restrictions on the geometry defined by restricting the holonomies of these connections.

\subsection{Generalised Holonomy in Generalised Geometry and Type I Extended Geometry}

As was seen in section \ref{gengen}, a type I extended tangent bundle is a bundle $E$ over a $d$-dimensional space $M$ with structure group $O(d,d)$ (or $SO(d,d)$), and reduces to a bundle $E^+\oplus E^-$
with structure group $O(d)\times O(d)$ or $S(O(d)\times O(d))$, and each sub-bundle is isomorphic to the tangent bundle, $E^\pm \sim T$ \cite{Gualtieri}. 

Consider first the case in which the extended geometry is a generalised geometry, which will be the case if the structure group of $E$ is in $GL(d,\R)\ltimes \W^{2,cl}$, so that for $E^+\oplus E^-$ it is in the diagonal $O(d)\subset O(d)\times O(d)$.
A generalised metric corresponds to a metric $G$ and closed 3-form $H$ on $M$, with 
$H=dB$ for some 2-form gerbe connection $B$.
Let $\nabla ^\pm$ be the metric connection on  $T$ given by the Levi-Civita connection plus torsion $\pm \frac 12 G^{-1}H$, $\nabla ^\pm = \nabla \pm \frac 12 G^{-1}H$, so that 
\begin{equation}
\nabla ^\pm _i v^j= \nabla _i v^j \pm \frac 12 H_{ik}^j v^k
\label{tors}
\end{equation}
where $H_{ik}^j = H_{ikl}G^{lj}$.

The holonomies of these connections, ${\cal H}^\pm =\ch (\nabla ^\pm)$, are in $O(d)$, ${\cal H}^\pm\subseteq O(d)$.
If $d=2m$ and ${\cal H}^+\subseteq U(m)$, then there is an almost complex structure $J^+$
that is parallel with respect to  $\nabla ^+$, $\nabla ^+J^+=0$.
 Similarly,  if ${\cal H}^-\subseteq U(m)$  there is an almost complex structure $J^-$
 with $\nabla ^-J^-=0$. The metric is hermitian with respect to each structure. An interesting case is that in which ${\cal H}^\pm \subseteq U(m)$, and this gives precisely the geometry needed to define a sigma-model with (2,2) world-sheet supersymmetry \cite{Gates:1984nk}.
 The superalgebra closes off-shell if both $J^\pm$ are integrable, and this gives precisely the bihermitian geometry of \cite{Gates:1984nk} which  has been termed generalised Kahler geometry in \cite{Gualtieri}. 

The isomorphism $E^\pm \sim T$ then gives corresponding connections $\nabla ^\pm$ on $E^\pm$, and the connection with supersymmetry suggests using the connection $\nabla ^+$ on $E^+ $ and the connection $\nabla ^-$ on $E^-$. Then the almost complex structures 
$J^\pm$ on $T$ correspond to generalised almost complex structures ${\cal J}_1, {\cal J}_2$
on $E$, and if    $J^\pm$ are integrable, then ${\cal J}_1, {\cal J}_2$ are Courant-integrable and so are 
generalised almost complex structures \cite{Gualtieri}.

There is a similar story for other holonomy groups \cite{Hull:1986iu},\cite{Hull:1986hn}. In table \ref{tab8}, 
the holonomy groups $\ch ^\pm$ that give sigma-models with $(p,q)$  
supersymmetry are given. (The cases $(q,p)$ are given by interchanging $\ch ^+,\ch^-$.)

 \begin{table}
\begin{center}
\begin{tabular}{|c|ccc|}
\hline
$(p,q)$ &$ \ch^+$ & $ \ch^-$ & dimension\\
\hline
$(1,1)$  & $O(d)$ &$ O(d) $& $d$ \\
$(2,1)$  & $U(m)$ &$ O(2m) $& $2m$ \\
$(2,2)$  & $U(m)$ &$ U(m)$& $2m$ \\
$(4,1)$  & $Sp(M)$ &$ O(4M) $& $4M$ \\
$(4,2)$  & $Sp(M)$ &$ U(2M) $& $4M$ \\
$(4,4)$  & $Sp(M)$ &$ Sp(M) $& $4M$ \\
\hline    
\end{tabular}
\caption{The holonomies $ \ch^+$, $ \ch^-$ giving $p-1$ complex structures $J^+$ and $q-1$ complex structures $J^-$ for manifolds of various dimension, which allow the construction of sigma-models with
$(p,q) $ supersymmetry. }
   \label{tab8} 
\end{center}
\end{table}

In each case, there are $p-1$ almost complex structures $J^+_\a$, $\a=1,..,p-1$ satisfying 
$\nabla ^+J^+_\a=0$, and $q-1$ almost complex structures $J^+_{\a'}$, $\a'=1,..,q-1$ satisfying 
$\nabla ^-J^-_{\a'}=0$. If there are three $J^+$ or $J^-$, they satisfy the quaternion algebra and so constitute an almost quaternionic structure.
Each pair $( J^+_\a, J^-_{\a'})$ defines two
generalised almost complex structures $\cj _1^{\a \a'}, \cj _2^{\a \a'}$ as in \cite{Gualtieri}, giving $2(p-1)(q-1)$
generalised almost complex structures.
For the $(4,2)$ case, there are 3+3 generalised almost complex structures $\cj _1^{\a }, \cj _2^{\a }$
satisfying an algebra with e.g.
\begin{equation}
[\cj _1^{\a }, \cj _1^{\b }]=[\cj _2^{\a }, \cj _2^{\b }]= \e ^{\a\b\g}\cj _1^{\g } \P ^+
\end{equation}
where $\P^\pm$ is the projection $\P^\pm : E \to E^\pm$.
For the $(4,4)$ case, there are 9+9 generalised almost complex structures $\cj _1^{\a  \a'}, \cj _2^{\a  \a'}$.
If all the almost complex structures are integrable, then the space is generalised Kahler if $p\ge2$ and $q\ge 2$. It seems natural to refer to the (4,4) case \cite{Gates:1984nk} as generalised hyperkahler, as  in \cite{Bredthauer:2006sz},\cite{Ezhuthachan:2006yy}.

The connections $\nabla ^\pm$ on $T$ lift to connections on the spin bundle (assuming $M$ is spin), with
\begin{equation}
\ti \nabla ^\pm _i \a= \nabla _i \a \pm \frac 18 H_{ijk}\G ^{jk} \a
\end{equation}
for spinors $\a$, where $\G^{ij}=\G^{[i}\G^{j]}$ and $\G^i$ satisfy the Clifford algebra
\begin{equation}
\{ \G^i , \G^j \} =2 G^{ij} \1
\end{equation}
The holonomies $\ti \ch ^\pm$   of $\ti \nabla ^\pm$ are in $Spin (d)$ and determine the number of covariantly constant spinors 
$\a^\pm$ satisfying $\ti \nabla ^\pm \a ^\pm=0$ \cite{Hull:1986iu}.
For general holonomy $\ti \ch ^+ = Spin(d)$, there are no covariantly constant spinors, while if
$d=2m$ and $\ti \ch ^+ \subseteq SU(m)$, then there are at least two satisfying $\ti \nabla ^+ \a ^+=0$.
The relation between holonomy and the number of parallel spinors is well-known:
for example, for $d=8$, there will be $1,2,3$ or $4$ such spinors for holonomies 
$Spin(7), SU(4), Sp(2), SU(2)\times SU(2)$ respectively, while for $d=7$, there is one such spinor for 
holonomy $G_2$.

Similar results apply for type I extended geometries.
A bundle $E$ with $O(d,d)$ structure   reduces to a bundle $E^+\oplus E^-$
with structure group $O(d)\times O(d)$.
In special cases, this will be reducible, and in this extended case, the structure group of $E^+$ need not be the same as that for $E^-$. 
The connections with torsion $\nabla ^\pm$ 
 again give connections on $E^\pm$, and we choose the connection $\nabla^+$ on $E^+$ and
$\nabla^-$ on $E^-$.
Again, there are interesting geometries with restrictions on the holonomies
$\ch ^\pm$.
For $d=2m$, bundles with $\ch^+\times \ch^-$ in $U(m)\times U(m)$ will be referred to as extended Kahler, and 
bundles with $\ch^+\times \ch^-$ in $SU(m)\times SU(m)$ will be referred to as extended Calabi-Yau.
The connections again lift to connections on the extended spin bundle with
structure $Spin(d)\times Spin(d)$, and the number of covariantly constant sections of these bundles play an important role in understanding supersymmetry in non-geometric backgrounds, as will be discussed elsewhere.

\subsection{Generalised Holonomy in Generalised Geometry and M-Extended Geometry}

For an M-geometry on an $n$-dimensional manifold $\cm$,  the extended tangent bundle $\ce$ has an $E_n$-structure and is reducible to one with compact structure group $H_n$, while the extended spin bundle $\ti \ce$  has structure $\ti H_n$.
For a conventional geometry, the structure groups reduce further to $SO(n)$ and $Spin(n)$ respectively, 
while the more general cases are relevant to non-geometric backgrounds.

Consider first the case of conventional geometry.
Sections of $\ti \ce $ are then spinor fields on $\cm$, and there is a natural connection on $\ti \ce $ that generalises (\ref{tors}), given by
\begin{equation}
{\tilde \nabla}_{i}= \nabla_{i}+ \frac{1}{24} \Gamma^{jkl}F_{ijkl}
  \label{supercovarianta}
\end{equation}
where  $F=dC$, $ \nabla_{i}$ is the usual spin connection,
$\Gamma_{i}$ are    Dirac matrices and $\Gamma_{ij...k}$ are  
antisymmetrised products of gamma matrices.
Note that, unlike (\ref{tors}), this does not project onto a connection on the tangent bundle.
Remarkably, this connection has holonomy $\ch$ that is always contained in $\ti H_n$ \cite{Hull:2003mf}. Interesting geometries arise when the holonomy is a special subgroup of $\ti H_n$.

This generalises to the case when the extended spin bundle is not reducible to the spin bundle, so that the structure group is in $\ti H_n$, and sections
are not spinor fields. The derivative (\ref{supercovarianta}) lifts to one acting on $\ti \ce$, and again the holonomy is in $\ti H_n$.

\subsection{Seven-Dimensional Spaces}

Consider the case in which ${\cal M}$ is  seven-dimensional, $n=7$.
For a Riemannian space with metric $G$, the holonomy $\ch(\nabla)$ of the Levi-Civita connection is in $SO(7)$, 
There will be at least one covariantly constant spinor satisfying $\nabla \a =0$ provided the holonomy is in $G_2$, $\ch(\nabla)\subseteq G_2$. 

For the extended spin bundle, the holonomy $\ch$  of the connection (\ref{supercovarianta}) is in $\ti H_7=SU(8)$.
There will be at least one section of $\ti \ce$ 
that is covariantly constant with respect to the connection (\ref{supercovarianta})  provided the holonomy is in the subgroup of $SU(8)$ preserving  an element $\a$  transforming in the {\bf 8} of $SU(8)$, 
$\ch \subseteq  U(7) \ltimes \C^7$.

\subsection{Relation with Supersymmetry}

For type I backgrounds, Killing spinors are spinors $\a^+, \a^-$ that are covariantly constant
\begin{equation}
\ti \nabla ^\pm \a ^\pm=0 
\end{equation}
and for which in addition there is a scalar $\Phi$ such that
\begin{equation}
\frac 16 H_{ijk} \G^{ijk} \a^\pm = \pm ( \pa _i \Phi )\G^i \a^\pm
\label{abc}
\end{equation}

The bosonic fields of 11-dimensional supergravity are a metric $G_{MN}$ and a 3-form gauge field $C_{MNP}$ ($M,N=0,1,...,10$),  with  a vielbein $e_M{}^A$ satisfying $e_M{}^A e_N{}^B\eta_{AB}=G_{MN}$ 
used to convert coordinate indices
$M,N$ to tangent space indices $A,B$.
The supercovariant derivative acting on spinors is
\begin{equation}
{\hat \nabla}_{M}= \nabla_{M}-
\frac{1}{288}(\Gamma_M{}^{NPQR}-8\delta_M^N\Gamma^{PQR})F_{NPQR},
\label{supercovariant}
\end{equation}
where $F=dC$,
the  $\Gamma_{A}$ are   $D=11$ Dirac matrices and $\Gamma_{AB...C}$ are  
antisymmetrised products of gamma matrices,  
$\Gamma_{AB...C}=\Gamma_{[A}\Gamma_{B}...\Gamma_{C]}
$. The signature is $(-++\dots+)$, and $\nabla_{M}$ is the
usual Riemannian covariant derivative involving the Levi-Civita
connection  
$\omega_{M}$
taking values in the tangent space  group $Spin(10,1)$
\begin{equation}
\nabla_{M}=\partial_{M}+\frac{1}{4}\omega_{M}{}^{AB}\Gamma_{AB}.
\end{equation}

Each solution of
\begin{equation}
{\hat \nabla}_{M}\epsilon=0,
  \label{supercovariantaa}
\end{equation}
is a Killing spinor field that
generates a supersymmetry leaving the background invariant, so that
the number of supersymmetries preserved by a supergravity background  
depends
on the number of supercovariantly constant spinors satisfying   
(\ref{supercovariantaa}).
Any commuting Killing spinor field $\epsilon$ defines a Killing vector  
$v_{A}=\overline{\epsilon}\Gamma_{A}\epsilon$, which is either timelike  
or null, together with a 2-form $\overline{\epsilon}\Gamma_{AB}\epsilon$
and a 5-form $\overline{\epsilon}\Gamma_{ABCDE}\epsilon$.

The integrability conditions for (\ref{supercovariantaa}) are satisfied if the background satisfies the supergravity field equations
 \begin{equation}
R_{MN}=\frac{1}{12}\left(F_{MPQR}F_{N}{}^{PQR}-\frac{1}{12}g_{MN}
F^{PQRS}F_{PQRS}\right)
\end{equation}
and
\begin{equation}
d*\!F+\frac 12F  \wedge F =0,
\end{equation}
 but the integrability conditions are weaker than the field equations.

Let
\begin{equation}
\label{fis}
f= \frac{1}{24}F_{MPQR} \Gamma^{MPQR}
\end{equation}
and note that the derivative (\ref{supercovariant}) can be rewritten as
\begin{equation}
{\tilde \nabla}_{M}= \nabla_{M}+ \frac{1}{24} \Gamma^{PQR}F_{MPQR}
- \frac{1}{12}\Gamma_Mf
  \label{supercovarianta}
\end{equation}
Then for backgrounds in which   the Killing spinor satisfies
\begin{equation}
\label{fiso}
f\epsilon  = 0\end{equation}
(such a constraint was used in \cite{Taylor},\cite{Beckera},\cite{Beckerb}, \cite{Hull:2003mf})
the Killing spinor condition simplifies to
\begin{equation}
{\ti \nabla}_{M}\epsilon \equiv (\nabla_{M}+ \frac{1}{24}  
\Gamma^{PQR}F_{MPQR})\epsilon=0
  \label{kil}
\end{equation}
and the   analysis of supersymmetric backgrounds in terms of the holonomy  
${\cal H}(\hat \nabla)$ \cite{Hull:2003mf}.

Consider product spaces ${\cal M}=M_{\ti n}\times M_{n}$ of spaces of dimensions $n, \ti n=11-n$,
so that the
coordinates can be split into $x^\mu, y^i$ with $\mu,\nu=1,...,\ti n=11-n $ and  
$i,j=1,...,n$, with a product
metric of the form
\begin{equation}
G_{MN}=
\left(\begin{array}{cc}
G_{\mu\nu}(x)&0\\
0&G_{ij}(y)
  \end{array}\right)
\end{equation}
where      $ g_{\mu\nu} (x)$ has Lorentzian signature  and $g_{ij}(y)$
has Euclidean signature. A convenient realisation of the gamma matrices
$\Gamma_M$ in terms of the gamma matrices $\gamma_\mu$ on $M_{\ti n}$ and the  
ones
$\ti \Gamma_i$ on $M_{n}$ is, for $n$ even,
\begin{equation}
\Gamma_\mu = \gamma_\mu \otimes  \ti  \Gamma_*, \qquad \Gamma_i = 1  
\otimes \ti  \Gamma_i
\end{equation}
where $\ti  \Gamma_*$ is the chirality operator on $M_{\ti d}$,  $\ti   
\Gamma_*\propto \prod_i  \ti  \Gamma_i $.
There is a similar realisation for $n$ odd.
  A spinor $\e$ decomposes as $\e=\h\otimes \a$ where $\h$ is a spinor on $M_{\ti n}$ and $\a$ is a spinor on $M_{n}$.
  
Suppose $M_{\ti n}$ is $\ti n$ dimensional Minkowski space with flat metric $G_{\mu\nu}$, and
  the only non-vanishing components of $F$ are $F_{ijkl}$ in the \lq internal space' $M_n$.
Then for any spinor $\a $ on $M_n$ satisfying
\begin{equation}
{\tilde \nabla}_{i}\a =0
  \label{supercovariantabc}
\end{equation}
where
\begin{equation}
{\tilde \nabla}_{i}= \nabla_{i}+ \frac{1}{24} \Gamma^{jkl}F_{ijkl}
  \label{supercovariantaaa}
\end{equation}
and the condition
\begin{equation}
F_{ijkl} \G^{ijkl} \a =0
\label{fghdgh}
\end{equation}
there will be a Killing spinor satisfying (\ref{supercovariantaa})  of the form $\h\otimes \a$ where $\h$ is any (covariantly) constant spinor in Minkowski space.
Thus supersymmetric backgrounds arise when the connection $\ti \nabla $ has a special holonomy so that there are solutions of (\ref{supercovariantabc}), and in addition each solution satisfies (\ref{fghdgh}).

\section*{Acknowledgements}

I would like to thank Dan Waldram and Nigel Hitchin for helpful  discussions. 
Seven-dimensional M-geometries with $E_7$ structure and their twisting by gerbes have been developed independently by Paulo Pires-Pacheco, Aaron Sim and Dan Waldram, as part of work in progress on 
flux compactifications to four dimensions.


\begin{thebibliography}{03}

\bibitem{Hitchin:2004ut}
N.~Hitchin,
Quart.\ J.\ Math.\ Oxford Ser.\  {\bf 54}, 281 (2003)
[arXiv:math.dg/0209099].

\bibitem{Hitchin:2005in}
  N.~Hitchin,
  arXiv:math.dg/0508618.

\bibitem{Hitchin:2005cv}
  N.~Hitchin,
  Commun.\ Math.\ Phys.\  {\bf 265}, 131 (2006)
  [arXiv:math.dg/0503432].

\bibitem{Gualtieri}
M.~Gualtieri,
Oxford University DPhil thesis, arXiv:math.DG/0401221.


\bibitem{Lindstrom:2004eh}
  U.~Lindstrom,
  Phys.\ Lett.\ B {\bf 587}, 216 (2004)
  [arXiv:hep-th/0401100].



\bibitem{glmw}S.~Gurrieri, J.~Louis, A.~Micu and D.~Waldram,
Nucl.\
Phys.\ B {\bf 654}, 61 (2003) [arXiv:hep-th/0211102].

 


\bibitem{Lindstrom:2005zr}
  U.~Lindstrom, M.~Rocek, R.~von Unge and M.~Zabzine,
  arXiv:hep-th/0512164.



\bibitem{Lindstrom:2004iw}
  U.~Lindstrom, R.~Minasian, A.~Tomasiello and M.~Zabzine,
  Commun.\ Math.\ Phys.\  {\bf 257}, 235 (2005)
  [arXiv:hep-th/0405085].
\bibitem{Grana:2004bg}
  M.~Grana, R.~Minasian, M.~Petrini and A.~Tomasiello,
  JHEP {\bf 0408}, 046 (2004)
  [arXiv:hep-th/0406137].
\bibitem{Grana:2004sv}
  M.~Grana, R.~Minasian, M.~Petrini and A.~Tomasiello,
  Comptes Rendus Physique {\bf 5}, 979 (2004)
  [arXiv:hep-th/0409176].
\bibitem{Grange:2004ah}
  P.~Grange and R.~Minasian,
  Nucl.\ Phys.\ B {\bf 732}, 366 (2006)
  [arXiv:hep-th/0412086].

\bibitem{Grana:2005sn}
  M.~Grana, R.~Minasian, M.~Petrini and A.~Tomasiello,
  JHEP {\bf 0511}, 020 (2005)
  [arXiv:hep-th/0505212].
\bibitem{Grange:2005nm}
  P.~Grange, R.~Minasian and E.~Polytechnique,
  Nucl.\ Phys.\ B {\bf 741}, 199 (2006)
  [arXiv:hep-th/0512185].
\bibitem{Minasian:2006hv}
  R.~Minasian, M.~Petrini and A.~Zaffaroni,
  arXiv:hep-th/0606257.


\bibitem{Zucchini:2004ta}
  R.~Zucchini,
  JHEP {\bf 0411} (2004) 045
  [arXiv:hep-th/0409181].
\bibitem{Lindstrom:2004cd}
  U.~Lindstrom,
  arXiv:hep-th/0409250.
\bibitem{Lindstrom:2004hi}
  U.~Lindstrom, M.~Rocek, R.~von Unge and M.~Zabzine,
  JHEP {\bf 0507} (2005) 067
  [arXiv:hep-th/0411186].

\bibitem{Jeschek:2004wy}
  C.~Jeschek and F.~Witt,
  JHEP {\bf 0503} (2005) 053
  [arXiv:hep-th/0412280].
\bibitem{Zucchini:2005rh}
  R.~Zucchini,
  JHEP {\bf 0503} (2005) 022
  [arXiv:hep-th/0501062].







\bibitem{Bredthauer:2006hf}
  A.~Bredthauer, U.~Lindstrom, J.~Persson and M.~Zabzine,
  arXiv:hep-th/0603130.



 

\bibitem{Cremmer:1979up}
  E.~Cremmer and B.~Julia,
  Nucl.\ Phys.\ B {\bf 159}, 141 (1979); B. Julia in {\it Supergravity and Superspace}, S.W. Hawking and M.
Ro$\check c$ek, C.U.P.
Cambridge,  (1981); B. Julia, {\it Infinite Lie algebras in Physics}
in {\it Johns Hopkins workshop on Unified field  theories
and beyond}, ed.  G. Domokos et al. Baltimore,  (June 1981).
  
\bibitem{Hull:1994ys}
C.~M.~Hull and P.~K.~Townsend,
Nucl.\ Phys.\ B {\bf 438}, 109 (1995) [arXiv:hep-th/9410167].


  
\bibitem{Keurentjes:2003hc}
  A.~Keurentjes,
  Class.\ Quant.\ Grav.\  {\bf 21} (2004) S1367
  [arXiv:hep-th/0312134].
\bibitem{Keurentjes:2003yu}
  A.~Keurentjes,
  Class.\ Quant.\ Grav.\  {\bf 21} (2004) 1695
  [arXiv:hep-th/0309106].
  


\bibitem{Hull:1997kt}
  C.~M.~Hull,
  Nucl.\ Phys.\ B {\bf 509} (1998) 216
  [arXiv:hep-th/9705162].

\bibitem{Cremmer:1998px}
  E.~Cremmer, B.~Julia, H.~Lu and C.~N.~Pope,
  Nucl.\ Phys.\ B {\bf 535}, 242 (1998)
  [arXiv:hep-th/9806106].


\bibitem{Hull:2004in}
  C.~M.~Hull,
JHEP {\bf 0510} (2005) 065,  arXiv:hep-th/0406102.

\bibitem{Hull:2006va}
  C.~M.~Hull,
  arXiv:hep-th/0605149.



\bibitem{Hull:2006qs}
  C.~M.~Hull,
  arXiv:hep-th/0604178.


\bibitem{Dabholkar:2002sy}
A.~Dabholkar and C.~Hull,
JHEP {\bf 0309}, 054 (2003) [arXiv:hep-th/0210209].



\bibitem{Dabholkar:2005ve}
  A.~Dabholkar and C.~Hull,
  JHEP {\bf 0605} (2006) 009
  [arXiv:hep-th/0512005].



\bibitem{Hull:2006tp}
  C.~M.~Hull and R.~A.~Reid-Edwards,
  arXiv:hep-th/0603094.


\bibitem{Flournoy:2004vn}
A.~Flournoy, B.~Wecht and B.~Williams,
arXiv:hep-th/0404217.

\bibitem{Kachru:2002sk}
S.~Kachru, M.~B.~Schulz, P.~K.~Tripathy and S.~P.~Trivedi,
JHEP {\bf 0303}, 061 (2003)
[arXiv:hep-th/0211182].

\bibitem{Hellerman:2002ax}
S.~Hellerman, J.~McGreevy and B.~Williams,
JHEP {\bf 0401}, 024 (2004)
[arXiv:hep-th/0208174].



\bibitem{Hull:2005hk}
  C.~M.~Hull and R.~A.~Reid-Edwards,
  arXiv:hep-th/0503114.


\bibitem{Flournoy:2005xe}
A.~Flournoy and B.~Williams,
arXiv:hep-th/0511126.

\bibitem{Shelton:2005cf}
  J.~Shelton, W.~Taylor and B.~Wecht,
  arXiv:hep-th/0508133.

\bibitem{Hull:2006qs}
  C.~M.~Hull,
  arXiv:hep-th/0604178.


\bibitem{Gray:2005ea}
  J.~Gray and E.~J.~Hackett-Jones,
  JHEP {\bf 0605} (2006) 071
  [arXiv:hep-th/0506092].


\bibitem{Lawrence:2006ma}
  A.~Lawrence, M.~B.~Schulz and B.~Wecht,
  JHEP {\bf 0607} (2006) 038
  [arXiv:hep-th/0602025].

\bibitem{Duff:2003ec}
  M.~J.~Duff and J.~T.~Liu,
  Nucl.\ Phys.\ B {\bf 674} (2003) 217
  [arXiv:hep-th/0303140].

\bibitem{Hull:2003mf}
  C.~Hull,
  arXiv:hep-th/0305039.



\bibitem{Hellerman:2006tx}
  S.~Hellerman and J.~Walcher,
  arXiv:hep-th/0604191.
  

  \bibitem{Hackett-Jones}
 Emily Hackett-Jones, George Moutsopoulos, arXiv:hep-th/0605114

\bibitem{Gray:2005ea}
  J.~Gray and E.~Hackett-Jones,
  arXiv:hep-th/0506092.



\bibitem{Gates:1984nk}
  S.~J.~Gates, C.~M.~Hull and M.~Rocek,
  Nucl.\ Phys.\ B {\bf 248} (1984) 157.




\bibitem{Hull:1986iu}
  C.~M.~Hull,
{\it  In  Proceedings of the 1st Torino Meeting on  Superunification and Extra Dimensions, 1985, World Scientific.}
%

\bibitem{Hull:1986hn}
  C.~M.~Hull,
{\it Lectures given at Super Field Theories Workshop, Vancouver, Canada, Jul 25 - Aug 6, 1986}



\bibitem{Karoubi}
M.~Karoubi.
\newblock Alg\`{e}bres de {C}lifford et {K}-th\'{e}orie.
\newblock {\em Ann. Scient. Ec. Norm. Sup.}, 1(1):161, 1968.


\bibitem{Hull:2000cf}
  C.~M.~Hull,
  JHEP {\bf 0006} (2000) 019
  [arXiv:hep-th/0004086].




\bibitem{Hull:2000zn}
  C.~M.~Hull,
  Nucl.\ Phys.\ B {\bf 583} (2000) 237
  [arXiv:hep-th/0004195].


\bibitem{Taylor}
S.~W.~Hawking and M.~M.~Taylor-Robinson,
Phys.\ Rev.\ D {\bf 58} (1998) 025006
[arXiv:hep-th/9711042].

\bibitem{Beckera}
K.~Becker and M.~Becker,
Nucl.\ Phys.\ B {\bf 477} (1996) 155
[arXiv:hep-th/9605053].

\bibitem{Beckerb}
K.~Becker,
JHEP {\bf 0105} (2001) 003
[arXiv:hep-th/0011114].

\bibitem{Bredthauer:2006sz}
  A.~Bredthauer,
  arXiv:hep-th/0608114.
\bibitem{Ezhuthachan:2006yy}
  B.~Ezhuthachan and D.~Ghoshal,
  arXiv:hep-th/0608132.




\end{thebibliography}
\end{document}